\documentclass[a4paper,12pt]{article}

\usepackage[english]{babel}  
\usepackage[utf8]{inputenc}

\usepackage{tcolorbox}
\usepackage{framed}
\usepackage{comment}
\usepackage{version}
\includeversion{ExtraComputation}

\usepackage{breakcites}
\usepackage{chngcntr}
\usepackage{tocloft}
\usepackage[totalwidth=16.5cm,totalheight=23.5cm]{geometry}
\usepackage{amsmath,amssymb,amsthm,mathtools}
 \usepackage{tikz}
 \usetikzlibrary{arrows,automata}
 \usetikzlibrary{positioning}

\definecolor{linkcolor}{rgb}{0,0,0.4} 
\usepackage[ pdftex,colorlinks=true,
pdfstartview=FitV,
linkcolor= linkcolor,
citecolor= linkcolor,
urlcolor= linkcolor,
hyperindex=true,
hyperfigures=false,
bookmarks=true]
{hyperref} 
\hypersetup{linktocpage}

\usepackage{bookmark}

\newtheorem{thmcnter}{thmcnter}[section]

\newtheorem{Proposition}[thmcnter]{Proposition}

\theoremstyle{definition}
\newtheorem{Definition}[thmcnter]{Definition}

\usepackage{bm}	
\newcommand{\Dt}{\tilde{D}}
\newcommand{\Xt}{\tilde{X}}

\newcommand{\ga}{\alpha}
\newcommand{\gb}{\beta}

\newcommand{\gc}{\gamma}
\newcommand{\gd}{\delta}
\newcommand{\gr}{\rho}
\newcommand{\eps}{\epsilon}
\newcommand{\epst}{\widetilde{\epsilon}}
\newcommand{\gt}{\tau}
\newcommand{\gtb}{\bar{\tau}}

\newcommand{\gl}{\lambda}

\newcommand{\gL}{\Lambda}
\newcommand{\gLt}{\widetilde{\Lambda}}
\newcommand{\gz}{\zeta}
\newcommand{\gzb}{\bar{\zeta}}
\newcommand{\gO}{\Omega}
\newcommand{\go}{\omega}
\newcommand{\gOb}{\overline{\Omega}}

\newcommand{\Psit}{\widetilde{\Psi}}
\newcommand{\gS}{\Sigma}
\newcommand{\gSt}{\widetilde{\Sigma}}
\newcommand{\bdgS}{\bm{\Sigma}}
\newcommand{\bdgSt}{\bm{\widetilde{\Sigma}}}
\newcommand{\gs}{\sigma}
\newcommand{\gsb}{\bar{\sigma}}
\newcommand{\gst}{\widetilde{\sigma}}

\newcommand{\J}{{\textit{J}}}

\newcommand{\At}{\widetilde{A}}
\newcommand{\bdA}{{\bm{A}}}
\newcommand{\bdAt}{{\bm{\widetilde{A}}}}
\newcommand{\bdB}{{\bm{B}}}
\newcommand{\Eb}{\bar{E}}
\newcommand{\bdf}{{\bm{f}}}	
\newcommand{\bdF}{{\bm{F}}}
\newcommand{\Ft}{\widetilde{F}}
\newcommand{\gti}{\tilde{g}}
\newcommand{\Gt}{\widetilde{G}}

\newcommand{\Mc}{\mathcal{M}}
\newcommand{\bdM}{{\bm{M}}}
\newcommand{\Oc}{\mathcal{O}}
\newcommand{\Tc}{\mathcal{T}}
\newcommand{\Wt}{\widetilde{W}}

\newcommand{\from}{\colon}
\newcommand{\inj}{\hookrightarrow}
\newcommand{\xto}[2][]{\xrightarrow[#1]{#2}}									
\newcommand{\R}{\mathbb{R}}
\newcommand{\C}{\mathbb{C}}
\newcommand{\Z}{\mathbb{Z}}
\newcommand{\CP}{\mathbb{C}\textbf{P}}
\newcommand{\W}{\wedge}
\newcommand{\N}{\nabla}
\newcommand{\Id}{\mathbb{Id}}
\newcommand{\pa}{\partial}
\newcommand{\id}{\lrcorner} 
\newcommand{\Ld}{\mathcal{L}} 
\newcommand{\su}{\mathfrak{su}}
\newcommand{\SO}{\rm{SO}}
\newcommand{\SU}{\rm{SU}}
\newcommand{\SL}{\rm{SL}}
\newcommand{\so}{\mathfrak{so}}

\newcommand{\Mtx}[1]{\begin{pmatrix} #1 \end{pmatrix}}

\newcommand{\T}{\mathbb{T}}

\newcommand{\PT}{\mathbb{PT}}
\newcommand{\PTc}{\mathcal{PT}}
\newcommand{\circonf }{\text{\textasciicircum}}
\newcommand{\pab}{\bar{\partial}}
\newcommand{\p}{\pi}						
\newcommand{\ph}{\hat{\pi}}	
\newcommand{\pp}{\pi.\hat{\pi}}

\begin{document}
 \pagestyle{plain}
\title{Pure Connection Formulation, Twistors and the Chase for a Twistor Action for General Relativity}
\author{Yannick Herfray \footnote{email adress: Yannick.Herfray@ens-Lyon.fr}\\ 
{\small \it Laboratoire de Physique, ENS de Lyon, 
	46 allée d'Italie, F-69364 Lyon Cedex 07, FRANCE} 
\\
{\small \it School of Mathematical Sciences, University of Nottingham, NG7 2RD, UK} }
\maketitle
\begin{abstract}\noindent
This paper establishes the relation between traditional results from (euclidean) twistor theory and chiral formulations of General Relativity (GR), especially the pure connection formulation. Starting from a $SU(2)$-connection only we show how to construct natural complex data on twistor space, mainly an almost Hermitian structure and a connection on some complex line bundle. Only when this almost Hermitian structure is integrable is the connection related to an anti-self-dual-Einstein metric and makes contact with the usual results. This leads to a new proof of the non-linear-graviton theorem. Finally we discuss what new strategies this `connection approach' to twistors suggests for constructing a twistor action for gravity. In appendix we also review all known chiral Lagrangians for GR.
\end{abstract}

\section*{Introduction}\pdfbookmark[section]{Introduction}{section: Introduction}
\counterwithout{equation}{section}	\setcounter{equation}{0}
\counterwithout{thmcnter}{section}	\setcounter{thmcnter}{0}

It is well known that general relativity (GR) can be given `chiral formulations' i.e formulations where the full local isometry group \begin{equation}\label{Introduction: SO4 decomposition}
	\SO(4,\C)= \SL(2,\C) \times \SL(2,\C)\big/ \Z^2 
\end{equation}
looses its central role for one of the `chiral' (left or right) subgroup $\SL(2,\C)$. For the most striking `chiral formulations' see \cite{Plebanski:1977zz}, \cite{Jacobson:1988yy}, \cite{Krasnov:2011pp}, see also appendix \ref{section: Appdx A Review of Chiral Lagrangians for Gravity} for a review of chiral Lagrangians for gravity. Note that the decomposition \eqref{Introduction: SO4 decomposition} applies to complexified gravity. This is for more generality as one can then restrict to any particular real case: in Lorentzian signature the two $\SL(2,\C)$ groups are complex conjugated while in Euclidean signature they are replaced by two independent $\SU(2)$ groups.

This shift in the local symmetries corresponds to a shift in the hierarchy of fields: in `chiral formulations' of GR the role of the metric is usually played down for other alternative variables with natural $\SL(2,\C)$ internal symmetries. Typically, the metric appears as a derived object and its associated local isometry group $\SO(4,\C)$ comes as an `auxiliary symmetry' that was somewhat hidden in the first place.

It's probably safe to say that the interest of the physics community for such reformulations started with Ashtekar `new' variables \cite{Ashtekar:1986yd} and the appealing form of the related diffeomorphism constraints.
In subsequent works \cite{Jacobson:1988yy}, \cite{Capovilla:1991qb} it was understood that the (ten year older!) Plebanski's action \cite{Plebanski:1977zz} gave a covariant description of Ashtekar variables. In Plebanski's pioneering work the metric completely disappears for $\SL(2,\C)$-valued fields. In both points of view, canonical and covariant, $\SL(2,\C)$-connections play a crucial role.

That $\SL(2,\C)$-connections appear is no surprise: In the more traditional metric perspective, the Levi-Civita connection comes as an $\SO(4,\C)$-connection. The decomposition of Lie group \eqref{Introduction: SO4 decomposition} then corresponds to a splitting of the Levi-Civita connection into Left(or self-dual) and Right(or anti-self-dual) $\SL(2,\C)$-connections, which are in some sense the most natural `chiral' objects one can construct from the metric. 
The `chiral formulations' of GR essentially reverse this construction: they take $\SL(2,\C)$ fields (e.g connections) as a building block for the metric. This culminates in the so called `pure connection of GR' pursued in \cite{Capovilla:1991kx} and finally achieved in \cite{Krasnov:2011pp} where the only field that appears in the Lagrangian is an $\SL(2,\C)$-connection. Chiral formulations of GR are reviewed from a general perspective in section \ref{section: Chiral formulation of gravity}, specific action principles are gathered in appendix \ref{section: Appdx A Review of Chiral Lagrangians for Gravity}.

On the other hand, it is not always apparent that twistor theory, at least in its original Penrose's program directed towards gravity\cite{Penrose:1999cw}, has a nice interplay with these reformulations and is in fact part of `chiral formulations' of GR in a broad sense. This is more clearly seen by taking a closer look at the main result of twistor theory on the gravity side, the `non-linear graviton theorem' \cite{Penrose:1976js}, \cite{Ward:1980am}.

The `non-linear graviton theorem' takes as a starting point an eight dimensional real manifold (the twistor space) equipped with an almost complex structure. This reduces the group of local symmetries to $\SL(4,\C)$ which is also the 4d (complex)conformal group $\SO(6,\C)  \simeq  \SL(4,\C)/\Z^2$ and indeed the first half of the non-linear graviton theorem asserts that, under some generic conditions, integrability of this almost complex structure is equivalent to a 4d complexified conformal anti-self-dual space-time (i.e such that self-dual part of Weyl curvature vanishes). That this theorem only describes anti-self-dual space-times clearly points in the direction of the intrinsic chirality of twistor theory, but there is more. 

The second half of the theorem requires additional data on twistor space in the form of a complex 1-form up to scale, usually denoted as $\gt$. This is essentially equivalent to a 4d real distribution at every point (the kernel of $\gt$). In order to describe space-times with non-zero cosmological constant, which is our main concerned in this paper, this distribution should also be taken to be non-integrable $\gt\W d\gt \neq 0$. This 1-form is taken to be `compatible' with the almost complex structure so that its kernel is in turn almost complex and identifies with $\C^2$. The restriction of the symmetry group $\SL(4,\C)$ to this distribution thus brings us down to the `chiral' group $\SL(2,\C)$: In fact such a 1-form is naturally associated with a `chiral' $\SL(2,\C)$-connection on space-time (This is especially clear in the Euclidean context and we will come back to this in what follows). As connections are not conformally invariant, it fixes a scale in the conformal space-time. The second part of the Non-Linear-Graviton Theorem then essentially asserts that this scale is such that the resulting metric is anti-self-dual Einstein if one is given a `good enough' (holomorphic) 1-form.

The usual approach to twistor theory generally emphasizes the metric aspect of the theorem and tend to overlook the fact that this 1-form, which crucially fixes the scaling to give Einstein equations, is directly related to a $\SL(2,\C)$-chiral connection thus putting twistor theory in the general framework of `chiral formulations of gravity'. In section \ref{section: Euclidean Twistor Theory Revisited} we will review the basics of the curved twistor construction with an emphasis on the relation between the chiral connection and the $\Oc(2)$-valued 1-form on twistor space $\gt$.

It is in fact well known to specialists that \emph{there is} an interplay between, for example, Plebanski formulation of GR and twistor theory as can be seen from the introduction of twistor variables in some recent spin-foam models \cite{Livine:2011vk,Speziale:2012nu} or in the conjoint use of Plebanski action and twistor theory \cite{Mason:2008jy} to investigate the structure of maximally helicity violating (MHV) gravity amplitudes. However, it is possible that not all consequences have been drawn from this overlapping.

Now, one of the `most radical' chiral formulations of GR is the pure connection formulation where only a $\SL(2,\C)$-connection is considered to be a fundamental field, the metric being a derived object. In this context Einstein equations take the form of second order field equations on the connection.

In this article, we wish to emphasize the change of perspective on twistor theory that this extreme chiral reformulation of gravity suggests: We already stated that, on twistor space, the equivalent of this chiral connection on space-time is a complex 1-form, $\gt$. In usual twistor theory this is just taken to be some additional data that complements the almost complex structure, the latter being fundamental. However the pure connection formulation of GR suggests that it is the 1-form $\gt$ ( loosely related again to the chiral connection) that should be taken as the starting point, with the almost complex structure (related the conformal structure) arising as a derived objects.

We demonstrate in section \ref{section: Euclidean Twistor Theory Revisited} that, at least in the Euclidean signature context, it is a valuable point of view and that it allows to reproduce nicely the results from the non-linear-graviton theorem while putting twistor theory firmly into the `chiral formulations' framework of gravity:

For a Riemannian manifold $M$ (i.e equipped with a metric of Euclidean signature) the associated twistor space $\T(M)$ is simply taken to be the 2-spinor bundle. Let $\left(\p_{A'},x\right)$ be coordinates adapted to the fibre bundle structure $\C^2 \inj \T(M) \to M$. Then an $SU(2)$-connection, $A^{A'}{}_{B'} \in \su(2)$, allows to define the 1-form on $\T(M)$:
\begin{equation*}
\gt = \p_{A'}\left(d\p^{A'} + A(x){}^{A'}{}_{B'} \p^{B'} \right)
\end{equation*}

related to the preceding discussion.

We first show that this is enough to construct a Hermitian structure on $\PT(M)$,  thus making contact with usual Euclidean twistor theory:
\begin{Proposition}{\emph{Almost Hermitian structure on $\PT(M)$}}\label{Proposition: introduction, AHS on PT}\mbox{} \\
	If $\bdA$ is a definite connection \emph{(see below for a clarification of this notion)} then $\PT(M)$ can be given an almost Hermitian structure, i.e a compatible triplet $\left(\J_A , \go_A , g_A \right)$ of almost complex structure, 2-form, and a Riemannian metric. 
	
	In general this triplet is neither Hermitian ($\J_A$ is not integrable) nor almost Kähler ($\go_A$ is non degenerate but generically not closed). In fact integrability of $\J_A$ is equivalent to the statement that $\bdA$ is the self-dual connection of a self-dual Einstein metric with non zero cosmological constant. The metric on twistor space can be made Kähler if and only if $\bdA$ is the self-dual connection of a self-dual Einstein metric with positive cosmological constant (i.e if the definite connection is of `positive sign' ).
	
	Further more, the integrability condition is equivalent to $\gt\W d\gt\W d\gt=0$.
\end{Proposition}
The main difference with the traditional results from \cite{Atiyah:1978wi} is that integrability is not only related to anti-self-duality but is irremediably linked to Einstein's equations. This is because in the construction described in \cite{Atiyah:1978wi} one is only interested in a conformal class of metrics while here the use of connections automatically fixes the `right scaling' that gives Einstein equations.

The fact that the connection needs to be `definite' refers to a natural non-degeneracy condition. Such connection can be assigned a sign. This terminology first appeared in \cite{Fine:2008} and we will review it in section \ref{section: Chiral formulation of gravity}.  
The possibility of associating a symplectic structure on $\PT(M)$ with a definite $SU(2)$-connection on $M$ was already pointed out in \cite{Fine:2008}. However, only in the integrable case does the symplectic structure described in this reference coincides with our $\go_A$. $\SL(2,\C)$-connections which are the self-dual connection of a self-dual Einstein metric with non zero cosmological constant were called `perfect' in \cite{Fine:2011} and are the one such that their curvature verify $F^i\W F^j \propto \gd^{ij}$. This well known (see e.g \cite{Capovilla:1990qi}) description of Einstein anti-self-dual metric in terms of connection will also be reviewed in section \ref{section: Chiral formulation of gravity}.
\\

On the other hand, starting with a certain 6D manifold $\PTc$, the projective twistor space, together with a 1-form valued in a certain line bundle $\gt$, we have a variant of the non linear graviton theorem:
\begin{Proposition}{\emph{Pure connection Non-Linear Graviton Theorem}} \label{Proposition: introduction, NLG} \mbox{}\\
	If $\gt$ is a definite 1-form then $\PTc$ can be given an almost complex structure $\J_{\gt}$.
	
	Together with some compatible conjugation operation on $\PTc$ this is enough to give $\PTc$ the structure of a fibre bundle over a 4d manifold $M$: $\CP^1 \inj \PTc \to M$.
	
	Integrability of $\J_{\gt}$ is then equivalent to the possibility of writing $\gt$ as 
\begin{equation*}
	\gt = \p_{A'}\left(d\p^{A'} + A^{A'}{}_{B'} \p^{B'} \right)	
\end{equation*}
	with $\bdA$ the self-dual connection of a Einstein anti-Self-Dual metric on $M$ with non zero cosmological constant.
	
	What is more the integrability condition reads $ \gt\W d\gt \W d\gt =0$.
\end{Proposition}

Bits and pieces of this last proposition were already known and developed in \cite{Mason05},\cite{Wolf:2007tx} and \cite{Adamo:2013tja} as part of a strategy to obtain twistor actions for conformal gravity, anti-self-dual gravity and gravity (the latter being still missing). However, we here give a new proof that emphasises the role of the connection as a fundamental object and we hope that by framing them in the general perspective of chiral approaches to gravity they will appear in a new light, i.e as more than just clever trick to construct twistor action.  In particular we hope to make it clear that one can effectively think of the (euclidean)non-linear-graviton theorem as a far reaching generalisation of the description of Einstein anti-self-dual metric in terms of connections. 

Our long term view in developing what could be called a `connection approach' to twistor theory, with the 1-form $\gt$ being the main field instead of the almost complex structure, was to open new strategies to construct twistor action for gravity. However one faces difficulties that we could not overcome. We briefly explain in section \ref{section: Discussion on the would be `Twistor action for Einstein gravity'} our work in this direction and why it does not seem to offer a way to a twistor action for gravity.\\

In this paper we consistently stick to the Euclidean signature. This is for coherence with our results concerning twistor theory which only apply to this signature. \\

This article is organised as follows: In section \ref{section: Chiral formulation of gravity} we review chiral formulations of gravity with an emphasis on the general geometric setting underlying any formulation of this type rather than on a particular Lagrangian. We especially stress how to write equations for self-dual gravity (i.e Einstein anti-self-dual metric) in this framework and review the pure connection field equations for Einstein metric. This will serve as a model for our `connection version' of the non-linear-graviton theorem. For completeness we gathered in appendix different Lagrangians that belong to the `chiral gravity' type, some of which might be unfamiliar to the reader. They will also be useful when considering the problem of a twistor action for gravity.

In section \ref{section: Euclidean Twistor Theory Revisited} we review some results from twistor theory in Euclidean signature but from an unusual connection perspective, i.e we take a $\SU(2)$-connection to be the main field instead of a metric. From this data only we show how to construct very natural structures on twistor space, namely the 1-form $\gt$, some associated connection on $\Oc(n)$ bundle and the triplet $\left(\J_A, \go_A, g_A\right)$ of compatible almost complex structure, 2-form and Euclidean metric on twistor space of proposition \ref{Proposition: introduction, AHS on PT}. We also review, from \cite{Fine:2008}, some symplectic structure that is naturally constructed from the connection. Finally we investigate the condition for integrability of the almost complex structure as well as the condition for which the triplet $\left(\J_A, \go_A, g_A\right)$ is Kähler. These cases turn out to be given by the self-dual-gravity equations and therefore make contact with the usual Kähler structure on twistor space constructed from an instanton (i.e an anti-self-dual Einstein metric).

We then state and give a new proof for the non linear graviton theorem from a pure connection point of view (cf proposition \ref{Proposition: introduction, NLG}). 

Finally in section \ref{section: Discussion on the would be `Twistor action for Einstein gravity'} we explain how ideas from sections \ref{section: Chiral formulation of gravity}-\ref{section: Euclidean Twistor Theory Revisited} suggest new ansätze for constructing twistor action for gravity. However this section will remain inconclusive and ideas described there should be seen as a few more elements on the chase (cf \cite{Mason05}, \cite{Adamo:2013tja}, \cite{Mason&Wolf09}, and \cite{Adamo:2013cra}) for this elusive (if existing) action.

\counterwithin{equation}{section}		\setcounter{equation}{0}
\counterwithin{thmcnter}{section}		\setcounter{thmcnter}{0}

\section{Chiral Formulations of Gravity} \label{section: Chiral formulation of gravity}

In this section we review `chiral formulations of gravity'. We try to adopt a broad perspective and describe the conceptual elements that are common to all these formulations rather than describe a particular Lagrangian. We also try to avoid hiding the simplicity of the geometrical concepts involved under a debauchery of indices. We however provide appendix \ref{section: Appdx Decomposition of the Curvature} for the reader interested in proofs in coordinates.

The main objective of this section is to introduce in a natural way the description of self-dual gravity in terms of $\SU(2)$-connections and the related notion of definite connections. This is achieved at the end of subsection \ref{ssection: Definite Connections and Gravity}. We also review how to write Einstein equations in terms of $\SU(2)$-connections only.

\subsection{Chiral Formulations of Gravity : Geometrical Foundations}\label{ssection: Chiral Formulations of GR - Fundations}

Chiral formulations of gravity exploit the fact that Einstein equations can be stated using only `one half' of the decomposition $\SO(4,\C) = \SL(2,\C) \times \SL(2,\C)/\Z^2$. We here briefly review why this is possible. 

The whole discussion in this section could be treated in complexified terms but for clarity and coherence with sections \ref{section: Euclidean Twistor Theory Revisited} and \ref{section: Discussion on the would be `Twistor action for Einstein gravity'} we will restrict to the real form $\SO(4,\R) = \SU(2) \times \SU(2)/\Z^2$, i.e Euclidean signature.

\subsubsection{Chiral decomposition of the curvature tensor}
Let us consider a Riemannian manifold $\left(M, g\right)$. We note $\left\{e^I\right\}_{I\in 0..3}$ an orthonormal co-frame and $\left\{e_I\right\}_{I\in 0..3}$ a dual frame, they are defined up to $\SO(4)$ transformations. In order to see that Einstein equations can be stated using only one half of the decomposition $\SO(4)=\SU(2) \times \SU(2)/\Z^2$, the quickest way is to split the Riemann curvature tensor into self-dual/anti-self-dual pieces. This is classically done in spinor notation (see e.g \cite{Penrose_vol1}) or more directly as in \cite{Atiyah:1978wi}. We here make a presentation along the line of the second reference with an emphasise on the necessity of using a torsion-free connection in order for chiral formulations of gravity to be possible. 
See also \cite{Capovilla:1991qb,Krasnov:2009pu} for pedagogical expositions.

As a starting point, let us consider a 2n-dimensional manifold. A crucial remark is that the hodge duality $* \from \gO^k(M) \to \gO^{2n-k}(M)$ sends n-forms on n-forms. Self-dual (resp anti-self-dual) n-forms are then eigenvectors with eigenvalues $+1$ (resp $-1$) for the hodge duality (In fact for a generic dimension and signature the eigenvalues are either $\pm1$ or $\pm i$ but we here already have in mind the application to Euclidean four dimensions). The case $2n=4$ is thus the only situation where 2-forms can be decomposed in self-dual $\gO^+$ and anti-seld-dual $\gO^+$ 2-forms. This happy accident has several implications.

By using the metric, 2-forms at a point $x \in M$ can be identified with anti-self-adjoint transformations of $\gO^1(M)_x$ and thus with $\so(2n)$:
\begin{equation}
b_{IJ} \frac{e^I \W e^J}{2} \in \gO^2(M)_x \quad \simeq \quad b_J{}^I \;e_I \otimes e^J\in End\left(\gO^1\right) \quad \simeq \quad \boldsymbol{b}\in \so(2n) 
\end{equation}
Where $b_{IJ} = b_I{}^K g_{KJ} $. The `accidental' split of Lie algebra 
\begin{equation}\label{Chiral Formulations of GR - Fundations: Lie algebra split}
\so(4)= \su(2)\oplus \su(2)
\end{equation}
then directly corresponds to the decomposition of 2-forms into self-dual and anti-self-dual 2-forms,
\begin{equation}\label{Chiral Formulations of GR - Fundations: 2-form split}
\gO^2 = \gO^2_+ \oplus \gO^2_-.
\end{equation}
Another accident is that \emph{curvature} forms are 2-forms. In four dimensions curvatures can thus be decomposed into smaller elementary bits. This is useful both for Yang-Mills type theories but also for gravity. We now turn to the decomposition of the Riemann tensor:

Consider a connection $\nabla$ on the tangent bundle compatible with the metric, this is a $\SO(4)$-connection (Note that, at this stage, we do not assume that the torsion of this connection vanishes). It splits into two $\SU(2)$-connections $D$ and $\widetilde{D}$,
\begin{equation}\label{Chiral Formulations of GR - Fundations: connection split}
\nabla = D + \widetilde{D}. 
\end{equation} 
They naturally act as connections on the bundle of self-dual 2-forms and anti-self-dual 2-forms respectively. 

As a consequence of \eqref{Chiral Formulations of GR - Fundations: connection split} the curvature $\N^2$ 2-form can be rewritten
\begin{equation}
\N^2 = D^2 + \Dt^2.
\end{equation}
At this point we only made use of the first `accident', the Lie algebra split \eqref{Chiral Formulations of GR - Fundations: Lie algebra split}. We can now make use of the second `accident' the 2-form decomposition \eqref{Chiral Formulations of GR - Fundations: 2-form split}: As we already pointed out, the curvature-forms $D^2$, $\Dt^2$ are indeed $\su(2)$-valued 2-forms or equivalently $\gO^+$ (resp $\gO^-$) -valued 2-forms,
\begin{equation}
D^2 \in \gO^2\left(M, \su(2)\right) \simeq \gO^2\left(M, \gO^2_+\right), \qquad \widetilde{D}^2 \in \gO^2\left(M, \su(2)\right) \simeq \gO^2\left( M,\gO^2_-\right).
\end{equation}

It follows from the decomposition, $\gO^2 = \gO^2_+ \oplus \gO^2_-$, that we can write them as bloc matrices:
\begin{equation}\label{Chiral Formulations of GR - Fundations: F, Ft decomposition}
D^2 = \Big( F , G \Big),\qquad \widetilde{D}^2= \left( \Gt , \Ft \right)
\end{equation}
where $F \in End\left(\gO^2_+\right)$, $G \in Hom\left(\gO^2_+, \gO^2_-\right)$, $\Ft \in End\left(\gO^2_-\right)$, $\Gt \in Hom\left(\gO^2_-, \gO^2_+\right)$.

Putting this altogether, the curvature of $\nabla$ can be written as a bloc matrix:
\begin{equation}\label{Chiral Formulations of GR - Fundations: Riemann decomposition 1}
\nabla^2 =
\left(\begin{array}{ll}
F & G \\ \Gt & \Ft
\end{array}\right) \quad \in \gO^2\left(M,\so(4)\right)\simeq End\left(\gO^2\right)
\end{equation}
It is also convenient to introduce the self-dual and anti-self-dual part of the Weyl tensor: 
\begin{equation}
\Psi = F -\frac{1}{3} tr F \; \Id, \qquad \Psit = \Ft -\frac{1}{3} tr \Ft \; \Id.
\end{equation}

Without any further assumptions this is as far as we can get. However, in the special case of the Levi-Civita connection, i.e if one assumes that the connection is torsion-free, we get a simpler picture: $\nabla^2\from \gO^2 \to \gO^2$ is then the usual Riemann curvature tensor and has some further symmetries. The torsion-free condition indeed implies that $\N^2$ has to be symmetric, i.e \begin{equation}\label{Chiral Formulations of GR - Fundations: Riemann symmetries}
G{}^t = \Gt,\qquad \Psi^t= \Psi,\qquad \Psit^t = \Psit.
\end{equation}
What is more, for the torsion-free connection $tr F = tr \Ft$. Using coordinates, one can indeed immediately see that this last identity is equivalent to the first Bianchi identity. See Appendix \ref{section: Appdx Decomposition of the Curvature} for more details.

From these considerations, we obtain the celebrated decomposition of the Riemann tensor into irreducible components:
\begin{equation}\label{Chiral Formulations of GR - Fundations: Riemann decomposition 2}
\nabla^2 = \underbrace{tr F \; \Mtx{ \Id & 0 \\ 0 & \Id}}_{\text{Scalar Curvature}}+ \underbrace{\Mtx{
		0 & G \\ G^t & 0
	}}_{\text{Ricci Traceless}} +\underbrace{\Mtx{
	\Psi & 0 \\ 0 & \Psit
}}_{\text{Weyl Curvature}}.
\end{equation}

Let us now turn to Einstein equations. From the above decomposition it stems that,
\begin{equation}\label{Chiral Formulations of GR - Fundations: Chiral Einstein Equations}
\text{g is Einstein if and only if} \quad G=0
\end{equation}
and then the scalar curvature is $4\gL = 4 tr F$.

In particular one sees from $D^2 = F + G$ that Einstein equations can be stated in term of $D$ only: The metric is Einstein if and only if $D$ is a self-dual gauge connection, i.e if $D^2$ is a self-dual $\su(2)$-valued 2-form.

Note however that, in order to achieve this `chiral formulation', the symmetries \eqref{Chiral Formulations of GR - Fundations: Riemann symmetries} were crucial. In case one does not assume the connection to be torsion-free the Riemann tensor does not enjoy the symmetries \eqref{Chiral Formulations of GR - Fundations: Riemann symmetries} and Einstein equations look much more complicated:
\begin{equation}\label{Chiral Formulations of GR - Fundations: Non-chiral Einstein equations}
G = -\Gt^t, \qquad \Psi^t = \Psi, \qquad \Psit^t = \Psit, \qquad \text{and}\quad \frac{tr F + tr \Ft }{2}= cst =\gL.
\end{equation}
See Appendix \ref{section: Appdx Decomposition of the Curvature} for a derivation in coordinates.

As opposed to \eqref{Chiral Formulations of GR - Fundations: Chiral Einstein Equations} this last set of equations involve the whole of the Riemann tensor and are therefore not `chiral' at all. One easily checks however that equations \eqref{Chiral Formulations of GR - Fundations: Non-chiral Einstein equations} together with the symmetries \eqref{Chiral Formulations of GR - Fundations: Riemann symmetries} give back the chiral formulation of Einstein equations \eqref{Chiral Formulations of GR - Fundations: Chiral Einstein Equations}, as it should.

From this presentation we hope to make it clear that the general phenomenon allowing for chiral formulations of Einstein equations stems from the internal symmetries of the Riemann tensor, related to torsion-freeness, and not from a particular choice of signature.

\subsubsection{Urbantke metric}\label{sssection: Urbantke metric}
In the above, we explained how Einstein equations can be stated in an essentially chiral way, i.e in terms of $\su(2)$-valued fields. This general principle underlies any chiral formulation of gravity. However this was still very classical in spirit as we considered the metric as the fundamental field. We now describe an essential observation due to Urbantke \cite{Urbantke:1984eb} that allows to obtain a metric as a derived object from chiral (i.e $su(2)$-valued) fields. 

Suppose that we have a $\su(2)$-valued 2-forms $\bdB$, using a basis of $\su(2)$, $\left(\gs_i\right)_{i\in 1,2,3}$, this gives us a triplet of real 2-form $\left(B^i\right)_{i \in 1,2,3}$, such that $\bdB= B^i\gs_i$

Now, given such a triplet of 2-forms $\left(B^i\right)_{i \in 1,2,3}$, there is a unique conformal structure that makes the triplet $\left(B^1,B^2,B^3\right)$ self-dual. We will refer to this conformal structure as the Urbantke metric associated with $\bdB$ and write it as $\tilde{g}_{\text{\tiny (B)}}$. There is even a way to make this conformal structure explicit through Urbantke formula \cite{Urbantke:1984eb},
\begin{equation}\label{Chiral Formulations of GR - Fundations: Urbantke metric}
\text{Urbanke metric}\text{:}\qquad \tilde{g}_{\text{\tiny (B)}}{}_{\mu\nu}\; = -\; \frac{1}{12}\epst^{\alpha\beta\gamma\delta}\eps_{ijk}B^i{}_{\mu\alpha}B^j{}_{\nu\beta}B^k{}_{\gamma\delta}.
\end{equation} 
Obviously, if the $B$'s do not span a 3 dimensional vector-space this cannot hold. In fact the `metric' \eqref{Chiral Formulations of GR - Fundations: Urbantke metric} will then be degenerated in the sense that it will not be invertible. A more precise statement, again from \cite{Urbantke:1984eb}, is the following:
given the triplet of two forms $\left(B^1,B^2,B^3\right)$, defines the conformal `internal metric' $\tilde{X}^{ij} = B^i \wedge B^j/ d^4x $ then Urbantke metric $\tilde{g}_{\text{\tiny (B)}}$ is invertible if and only if $\tilde{X}$ is. When Urbantke metric is invertible $\tilde{X}$ is just the metric on the space of self-dual 2-forms given by wedge product.

As we started with a triplet $\left(B^i\right)_{i\in 1,2,3}$ of \emph{real} 2-forms, the associated Urbantke metric \eqref{Chiral Formulations of GR - Fundations: Urbantke metric} is also real. One the other hand, \emph{its signature is undefined}: self-dual 2-forms in Lorentzian signature are complex so this signature is excluded but without further restriction it can still be either Euclidean or Kleinian. The signature of the internal metric $\tilde{X}$ however is enough information to fix this ambiguity: for an Euclidean conformal metric $\tilde{g}$ the metric $\tilde{X}$ on self-dual 2-forms given by wedge product is Euclidean while for a Kleinian signature it would be Lorentzian.

Thus if we start with a triplet $\left(B^i\right)_{i\in 1..3}$ of real 2-form such that the internal metric $\tilde{X}^{ij} = B^i\W B^j \big/ d^4x$ is definite, we are then assured that the associated Urbantke metric, $\tilde{g}_{\text{\tiny (B)}}$ is non degenerate (invertible) and of Euclidean signature. This suggests to introduce the following definition:
\begin{Definition}{\emph{Definite Triplet of 2-forms}}\label{Definition- Chiral Formulations of GR - Fundations: definite triplet} \mbox{} \\
	A triplet $\left(B^1, B^2, B^3\right)$ of real 2-forms is called definite if the conformal metric constructed from the wedge product $\tilde{X}^{ij} = B^i\W B^j\big/ d^4x$ is definite.
\end{Definition}
As we just explained this is a useful definition because of the following:
\begin{Proposition}{\emph{Urbantke metric}} \mbox{} \label{Proposition- Chiral Formulations of GR - Fundations: Urbantke metric}\\
	The Urbantke metric \eqref{Proposition- Chiral Formulations of GR - Fundations: Urbantke metric} associated with a definite triplet of 2-forms is non degenerate and of Euclidean signature.
\end{Proposition}
\begin{proof}
	The equivalence between the non-degeneracy of $\gti_{\text{\tiny (B)}}{}_{\mu\nu}$ and the non-degeneracy of $\Xt^{ij}$ can be found in \cite{Urbantke:1984eb}. This reference also shows that the $B$'s are self-dual for $\gti_{\text{\tiny (B)}}{}_{\mu\nu}$. As already discussed above the 3D wedge product metric on self-dual 2-forms $\Xt$ can only be real definite for a (conformal) Euclidean four dimensional metric.
\end{proof}
In this section we made two distinct but complementary observations, first Einstein equations can be stated in a chiral way (cf equation \eqref{Chiral Formulations of GR - Fundations: Chiral Einstein Equations}) ie using $\su(2)$-valued fields, second a (definite) $\su(2)$-valued 2-form is enough to define a metric. Lagrangians that realise `chiral formulations' of GR all rely on some mixture of these two facts each with its own flavour and fields. See appendix \ref{section: Appdx A Review of Chiral Lagrangians for Gravity} for explicit variational principles.

However, for the most of this part, we won't be interested by a particular action but rather by how the general framework that we just describe intersects with twistor theory. Our main guide will be the description of anti-self-dual Einstein metric in terms of connections. 

Before we come to this it is useful to introduce two new tensors.

\subsubsection{Two useful tensors: the sigma 2-forms}\label{sssection: Two useful tensors}

We already made the remark that a metric allows to identify the Lie algebra $\so(4)$ with the space of 2-forms $\gO^2$. We denote by  \begin{equation}
\Phi \from \so(4)= \su(2)\oplus\su(2) \to \gO^2=\gO^2_+\oplus \gO^2_-
\end{equation} this isomorphism.

We choose a basis $\left( \,\gs^i , \gst^i \right)_{i\in 1,2,3}$ of $\so(4)= \su(2)\oplus\su(2)$ adapted to the decomposition and such that 
\begin{equation}
\left[\gs^i, \gs^j\right] = \eps^{ijk}\,\gs^k  ,\quad  \left[\gst^i, \gst^j\right] = \eps^{ijk}\gst^k ,\quad \left[\gs^i, \gst^j\right] = 0.
\end{equation}
Then one can define the sigma 2-forms:
\begin{equation}\label{Chiral Formulations of GR - Fundations: Sigma def (geometric)}
\frac{1}{2} \gS^i = \Phi \left( \gs^i \right), \qquad \frac{1}{2} \gSt^i = \Phi \left( \gst^i \right).
\end{equation}
Thus $\Big(\gS^i \Big)_{i\in 1,2,3}$ (resp $\left(\gSt^i \right)_{i\in 1,2,3}$) form a basis of self-dual 2-forms $\gO^2_+$ (resp anti-self-dual 2-forms $\gO^2_-$). This basis is also defined (up to $SU(2)$ transformations) by the orthogonality relations
\begin{equation}
\gS^i \W \gS^j = -\gSt^i \W \gSt^j = 2 \gd^{ij} Vol_g, \qquad \gS^i \W \gSt^j = 0.
\end{equation}

The awkward factor of one half in the definition is there for it to fit with the definition in terms of a tetrad that frequently appears in the literature:
\begin{align}\label{Chiral Formulations of GR - Fundations: Sigma def (tetrad)}
&\left(\gS^i= -e^0 \W e^i - \frac{\eps^{ijk}}{2} e^j \W e^k\right)_{i\in 1,2,3}, \qquad  
&\left(\gSt^i = e^0 \W e^i - \frac{\eps^{ijk}}{2} e^j \W e^k\right)_{i\in 1,2,3}.
\end{align}

The sigma 2-forms are naturally $\su(2)^*$-valued 2-forms or, using the Killing metric on $\su(2)$, $\su(2)$-valued 2-forms:
\begin{equation}
\bdgS = \gS^i \,\gs^i \in \gO^2_+\left(M,\su(2)\right), \qquad \bdgSt =\gSt^i \,\gst^i \in \gO^2_-\left(M,\su(2)\right).
\end{equation}
 (In this article bold notation will indicate $\su(2)$-valued objects)
 
Importantly they are compatible with the connections $D= d + \bdA$ , $\Dt = d + \bdAt$, in the following sense:
\begin{Proposition}\mbox{}\\
	Let $\bdgS$ (resp $\bdgSt$) be the $\su(2)$-valued self-dual (resp anti-self-dual) 2-forms constructed from a metric as \eqref{Chiral Formulations of GR - Fundations: Sigma def (tetrad)}, let $D= d + \bdA$ be the self-dual part of the Levi-Civita connection associated with this metric. Then $D$ is the $\SU(2)$ connection satisfying
	\begin{equation}\label{Chiral Formulations of GR - Fundations: Sigma/D compatibility}
	d_{\bdA} \Big(\bdgS \Big)= d\bdgS + [\bdA , \bdgS ]  =0, \qquad d_{\bdAt} \left(\bdgSt \right)= d\bdgSt + [\bdAt , \bdgSt ]   =0.
	\end{equation}
\end{Proposition}
\begin{proof}
	See Appendix \ref{section: Appdx Decomposition of the Curvature} for a direct proof in coordinates.
\end{proof}
This compatibility relations are important as they can be used as alternative definition for the chiral connection $D$ and $\Dt$.

Finally we can write the Einstein equations in terms of those 2-forms. If $D=d +\bdA$ is the `left' or `self-dual' connection and $D^2 = \bdF$ its curvature, then we can rewrite the first half of the bloc decomposition \eqref{Chiral Formulations of GR - Fundations: Riemann decomposition 2} as
\begin{equation}\label{Chiral Formulations of GR - Fundations: F, Ft decomposition 2}
D^2 = F^i\,\gs^i = \left( F^{ij} \gS^j + G^{ij}\gSt^j \right)\gs^i.
\end{equation}
Then, as we already discussed, the self-dual part of Weyl curvature is \begin{equation}
\Psi^{ij} =F^{ij} - \frac{1}{3}tr F \gd^{ij},
\end{equation} the scalar curvature is $4 \gL= 4tr F$ and
\begin{equation}\label{Chiral Formulations of GR - Fundations: Chiral Einstein Equations 2}
\text{g is Einstein if and only if} \quad D^2 = M^{ij} \gS^j \,\gs^i.
\end{equation}

\subsection{Definite Connections and Gravity} \label{ssection: Definite Connections and Gravity}

We review here how to write equations for Einstein-anti-self-dual metric in terms of connections. This is a well known construction (cf  \cite{Capovilla:1990qi}) and we here use the terminology of \cite{Fine:2008},\cite{Fine:2011}. We also briefly recall how to write equations for full Einstein gravity in terms of connections from \cite{Krasnov:2011pp},\cite{Fine:2013qta}.

We now take $\bdA = A^i \,\gs^i$ to be the potential in a trivialisation of a $SU(2)$-connection $D =d + \bdA$ and $D^2 = \bdF = F^i \,\gs^i $ its curvature.

\subsubsection{Definite Connections}\label{sssection: Definite Connections}

We mainly consider definite connections, i.e connections such that the curvature 2-form is a definite triplet:
\begin{Definition}{\emph{Definite Connections}}\label{Definition- Definite Connections and Gravity: Definite Connections}\mbox{} \\
	A $SU(2)$-connection $D = d + A^i \,\gs^i$, is called \emph{definite} if the conformal metric, $\tilde{X}^{ij} = F^i\W F^j \big/d^4x$, constructed from its curvature, $D^2=F^i \,\gs^i$, is definite. 
\end{Definition}
For any $SU(2)$-connection with potential $A^i \,\gs^i$, there is a unique conformal class of metric $\tilde{g}_{\text{\tiny(F)}}$ such that the curvature $F^i \,\gs^i$ is self-dual. The definiteness of the connection then ensures that this conformal metric is invertible and of Euclidean signature (cf Def \ref{Definition- Chiral Formulations of GR - Fundations: definite triplet} and Prop \ref{Proposition- Chiral Formulations of GR - Fundations: Urbantke metric} ). Thus definite connections are associated with a `good' metric.

A definite connections also defines a notion of orientation. It is done by restricting to volume form $\nu_+$ such that $\tilde{X}^{ij} = F^i \W F^j \big /\nu_+ $ is \emph{positive} definite. In the following whenever there is a need for an orientation, we will always take this one.

We can also assign a \emph{sign} to a connection as follows: We consider co-frame $\left(e^I\right)_{I\in 0..3}$, orthonormal with respect to the Urbantke metric and oriented with the convention that we just described. They are defined up to Lorentz transformations and rescaling by a \emph{positive} function. From this tetrad we can construct a basis of self-dual 2-form $\left(\gS^i \right)_{i\in 1,2,3}$ through the relation \eqref{Chiral Formulations of GR - Fundations: Sigma def (tetrad)}. Again $\left(\gS^i\right)_{i\in 1,2,3}$ is defined up to $SU(2)$ transformations and rescalings by \emph{positive} functions.
By construction, the curvature $D^2 =F^i \,\gs^i$ is self-dual for the associated Urbantke metric and we can thus write
\begin{equation}
D^2 = F^i \,\gs^i = M^{ij} \gS^j \,\gs^i.
\end{equation}
The sign of the connection is then defined as $s = \text{sign}\left(det\left(M\right)\right)$.
Note that this notion of sign makes sense as a result of $\left\{F^i\right\}_{i\in1,2,3}$ being defined up to $SU(2)$ transformations and $\left\{\gS^i\right\}_{i\in1,2,3}$ being defined up to $SU(2)$ transformations and positive rescaling.

We now have two $SU(2)$ transformations independently acting on $\left(F^i\right)_{i\in1,2,3}$ and $\left(\gS^i\right)_{i\in1,2,3}$, the first as a result of changing the trivialisation of the $SU(2)$ principal bundle of whom $D= d+ \bdA$ is a connection, the second as a result of changing the trivialisation of the bundle of self-dual 2-forms associated with the Urbantke metric. Those two bundles can be identified (at least locally) by requiring $M^{ij}$ to be a definite symmetric matrix. Finally we also have two scaling transformations, one acting on $\Xt$ and the other one on $\gS$, we identify them by requiring that $ F^i \W F^j = \Xt^{ij} \; \frac{1}{3} \gS ^k \W \gS^k$.

In what follows these identifications will always be assumed unless we explicitly specify otherwise.

As a result of $\Xt^{ij}$ being definite we can make sense of its square root. In fact there is a slight ambiguity in this definition: we fix it by requiring $\sqrt{X}$ to be positive definite, i.e we take the positive square root.

With these choices of square root and identifications, we have
\begin{equation}
F^i= s\, \sqrt{X}^{ij}\gS^j \qquad \Leftrightarrow \qquad \gS^i = s \, \sqrt{X}^{-1}{}^{ij} F^j, \qquad \forall i \in 1,2,3.
\end{equation}

\subsubsection{Anti-self-dual gravity and Perfect Connections}\label{sssection: Self-dual gravity and Perfect Connections}

A metric is said to be `anti-self-dual' if the self-dual part of its Weyl curvature vanishes ie, if $W_+=0$ in \eqref{Chiral Formulations of GR - Fundations: Riemann decomposition 2}. 
As Weyl curvature is conformally invariant, this is a property of the conformal class of the metric rather than from the metric itself. 

A metric is Einstein-anti-self-dual if it is Einstein and anti-self-dual, ie if $W_+=0$ , $G=0$ in \eqref{Chiral Formulations of GR - Fundations: Riemann decomposition 2}. Alternatively, using \eqref{Chiral Formulations of GR - Fundations: F, Ft decomposition 2}, if
\begin{equation}\label{Definite Connections and Gravity: ASD metric def}
F^i = \frac{\gL}{3} \gS^i , \qquad \forall i \in 1,2,3.
\end{equation}
(then $4\gL$ is the scalar curvature). Note in particular that for $\gL\neq0$, $F^i\W F^j \big/ d^4x \propto \gd^{ij}$.

This motivates the following definition,
\begin{Definition}{\emph{Perfect Connections}}\mbox{}\\
	A definite connection is perfect if $F^i\W F^j = \gd^{ij} \frac{F^k \W F^k}{3}$.
\end{Definition}

The relevance of this definition comes from the following:
\begin{Proposition}\label{Proposition- Definite Connections and Gravity: Perfect connection}\mbox{}\\
	The Urbantke conformal metric associated with a perfect connection is anti-self-dual. What is more the representative with volume form $ \frac{3}{2\gL^2} F^k \W F^k$ is anti-self-dual-Einstein with cosmological constant $s \,|\gL|$, where $s$ is the sign of the connection.
\end{Proposition}

\begin{proof}\mbox{} \\
	Consider the Urbantke metric with volume form $\nu = \frac{3}{2 \gL^2} F^k \W F^k$. It is associated with a orthonormal basis of 2-form $\left\{\gS^i\right\}_{i\in1,2,3}$ as in \eqref{Chiral Formulations of GR - Fundations: Sigma def (tetrad)}. By construction, they are such that $\gS^i \W \gS^j = 2 \gd^{ij} \; \nu$. 
	Together with our identification of the scaling transformations, $ F^i\W F^j = \Xt^{ij} \frac{1}{3} \gS^k \W \gS^k$, it gives
	 \begin{equation}
	 F^i\W F^j = 2\Xt^{ij} \nu.
	 \end{equation}
	Now by hypotheses,
	 \begin{equation}
	 F^i \W F^j = \frac{\gd^{ij}}{3} F^k \W F^k = 2\gd^{ij} \frac{\gL^2}{9} \nu,
	 \end{equation}
	from which we read $X^{ij} = \frac{\gL^2}{9} \gd^{ij}$ and
	\begin{equation}\label{Definite Connections and Gravity: Proof, SD gr eq1}
	F^i = s \, \sqrt{X}^{ij}\gS^j = s \, \frac{|\gL|}{3} \gS^i.
	\end{equation}
	From this last relation we see that Bianchi identity, $d_{\bdA} \bdF=0$, is now equivalent to $d_{\bdA} \bdgS= d\bdgS + [\bdA, \bdgS] =0$ which is the defining equation \eqref{Chiral Formulations of GR - Fundations: Sigma/D compatibility} of the self-dual connection. It follows that $D= d + \bdA$ is the self-dual connection of the Urbantke metric with volume form $\nu =\frac{3}{2 \gL^2} F^k \W F^k$. With this observation \eqref{Definite Connections and Gravity: Proof, SD gr eq1} are just the field equations for Einstein anti-self-dual gravity \eqref{Definite Connections and Gravity: ASD metric def} with cosmological constant $s\,|\gL|$.
\end{proof}

\subsubsection{Pure connection formulation of Einstein equations}\label{sssection: Pure connection formulation of Einstein equations} 
At this point it is hard to resist writing down the pure connection formulation of Einstein equations.\\

Consider a definite $SU(2)$-connection $D=d + \bdA$ with curvature $\bdF=F^i\,\gs^i$. As already explained, it is associated with an orientation, a sign $s$ and conformal class of metric $\tilde{g}_{\text{\tiny(F)}}$. We again denote $F^i\W F^j = \tilde{X}^{ij} d^4x$ and define the following volume form,
\begin{equation}
\frac{1}{2\gL^2}\left(Tr\sqrt{F\W F}\right)^2 \coloneqq \frac{1}{2\gL^2}\left(Tr\sqrt{\tilde{X}}\right)^2 d^4x .
\end{equation}
This is a well defined expression as a result of the following facts: the definiteness of the connection together with the orientation make $\tilde{X}^{ij}$ positive definite and thus we can take its square root, what is more $\left(Tr\sqrt{\tilde{X}}\right)^2$ being homogeneous degree one in $\tilde{X}$ the overall expression does not depends on the representative of the density $\tilde{X}$.

However there are signs ambiguity in this choice of square root. They amount to the choice of signature of the conformal metric $\sqrt{X}^{ij}$. We will always take this choice of square root such that $det \left(\sqrt{X}\right) >0$, then the only signatures that remains are $(+,+,+)$ and $(+,-,-)$. We thus need to make a choice for our definition of square root once and for all: either we stick with the `definite square root' or with the `indefinite square root'. 

\begin{Definition}{\emph{Einstein Connections}}\label{Definition- Definite Connections and Gravity: Einstein connection} \mbox{}\\
	If $A^i$ is a definite connection, define $X^{ij}$ by the relation
	\begin{equation}\label{Definite Connections and Gravity: Proof, pure connection eq1}
	F^i \W F^j = 2 X^{ij}\frac{1}{2\gL^2}\left(Tr\sqrt{F\W F}\right)^2 .
	\end{equation}
	Then we will call it Einstein if
	\begin{equation}\label{Definite Connections and Gravity: Pure connection Einstein equations}
	d_A \left( \left(\sqrt{X}\right)^{-1}{}^{ij} F^j \right)=0.
	\end{equation}
\end{Definition}
Again, the two square roots in this definition need to be taken with the same convention, ie such that the resulting matrices have the same signature: either $(+,+,+)$ ( `definite square root') or indefinite $(+,-,-)$ (`indefinite square root'). Note that for perfect connections, $X^{ij} = \gd^{ij}\frac{\gL^2}{9}$, as a result of which perfect connections are special case of Einstein connections with the `definite square root' convention (note that perfect connections are \emph{not} Einstein connections for the `indefinite square root' as $d_A \left( \left(\sqrt{X}\right)^{-1}{}^{ij} F^j \right) \neq 0$ for $\sqrt{X}= diag\left(1,-1,-1\right)$).

The Definition \ref{Definition- Definite Connections and Gravity: Einstein connection} is motivated by the following,
\begin{Proposition}{\emph{ Krasnov \cite{Krasnov:2011pp}}}\label{Proposition- Definite Connections and Gravity: Pure connection equation}\mbox{}\\
	For an Einstein connection, the Urbantke metric with volume form $\frac{1}{2\gL^2}\left(Tr\sqrt{F \W F}\right)^2$ is Einstein with cosmological constant $|\gL| \text{Sign}\left(s \, Tr\sqrt{F \W F} \right) $. What is more such a connection coincides with the self-dual Levi-Civita connection of the metric.
\end{Proposition}

\begin{proof} \mbox{}\\
	The metric in Urbantke conformal class with volume form $\nu=\frac{1}{2\gL^2}\left(Tr\sqrt{F \W F}\right)^2$ is associated with an orthonormal basis of self-dual 2-form $\left\{\gS^i\right\}_{i\in1,2,3}$, $\gS^i\W \gS^j = 2 \gd^{ij}\nu$. It is defined up to $\SU(2)$ transformation. By definition, $\left\{F^i\right\}_{i\in1,2,3}$ is a basis of self-dual 2-forms for Urbantke metric and $F^i = M^{ij} \gS^j \; \forall i\in \{1,2,3\}$.
	
	As was already pointed out, \emph{a priori} $\bdF$ and $\bdgS$ are valued in two different associated $\SU(2)$ bundle: $D=d+\bdA$ is a $\SU(2)$-connection on a $\SU(2)$ principal bundle P and the curvature naturally takes value in the adjoint bundle $ P \times_{SU(2)} \su(2)  $,  on the other hand $\left\{\gS^i\right\}_{i\in1,2,3}$ is a trivialisation of the bundle of self-dual 2-forms associated with the Urbantke metric.
	
	We now come again to the subtle question of identifying the two: this can be done (at least locally) by requiring $M^{ij}$ to be a symmetric matrix. Once this is done, however there is still the possibility of acting with the diagonal transformation $\left(\gS^1, \gS^2, \gS^3 \right) \to \left(\gS^1, -\gS^2, -\gS^3 \right)$ and we thus have two possible identifications. We call them the `definite identification' and the `indefinite identification' depending whether or not the resulting matrix $M^{ij}$ is definite or not.
	
	As a rule, we now take the identification corresponding to the square root that we chose, ie if one chooses the `definite square root', we take the `definite identification'; on the other hand, if one takes the `indefinite square root' one should use the `indefinite identification'.
	
	Finally, just as in the case of perfect connections, we identify rescaling of $\Xt$ and rescaling of $\gS$ by imposing that $F^i\W F^j =  \Xt^{ij} \;\frac{1}{3} \gS^k \W \gS^k$. Together with the choice of volume form, $\gS^i \W \gS^j = 2 \gd^{ij} v$, this completely fixes all the scaling freedom: $F^i \W F^j = 2 X^{ij} \nu$. Note that this gives the same result as in definition \ref{Definition- Definite Connections and Gravity: Einstein connection}.
	
	As a consequence of these different choices we have
	\begin{equation}
	F^i= s\, \sqrt{X}^{ij}\gS^j \qquad \Leftrightarrow \qquad \gS^i =  s\, \sqrt{X}^{-1}{}^{ij} F^j.
	\end{equation}
	The field equations \eqref{Definite Connections and Gravity: Pure connection Einstein equations} now read $d_{\bdA} \bdgS=0$ which are just the the defining equations \eqref{Chiral Formulations of GR - Fundations: Sigma/D compatibility} of the self-dual connection. It follows that $D=d + \bdA$ is the self-dual connection of the Urbantke metric with volume form $\nu$. Having this in mind, $F^i= s\, \sqrt{X}^{ij}\gS^j$, are Einstein equations \eqref{Chiral Formulations of GR - Fundations: Chiral Einstein Equations 2} with cosmological constant $s\,Tr\left(\sqrt{X}\right)$ . Finally, from \eqref{Definite Connections and Gravity: Proof, pure connection eq1}, one gets $|Tr\left(\sqrt{X}\right)| = |\gL|$.
\end{proof}

Note that one of the weakness of this formulation is that a particular choices of square root (ie `definite' or `indefinite') can only describe a particular subspace of Einstein metric, those such that the self-dual Weyl curvature $F^{ij}$ is respectively definite or indefinite. 

Interestingly, the integral of the volume form $\frac{1}{2\gL^2}\left(Tr\sqrt{F \W F}\right)^2$ also gives the correct variational principle for Einstein connections. This is the \emph{pure connection action} for GR \cite{Krasnov:2011pp}. It can be obtained by integrating fields successively from the Plebanski action, see appendix \ref{section: Appdx A Review of Chiral Lagrangians for Gravity}.

\section{Euclidean Twistor Theory Revisited: a Connection Point of View}  \label{section: Euclidean Twistor Theory Revisited}

We now review the Euclidean version of the twistor construction but from an unusual `connection point of view'.\\

We take `space-time' as a $\SU(2)$-principal bundle
 \begin{equation}
\SU(2) \inj P \to M
\end{equation}
 over a 4d manifold $M$ equipped with a $\SU(2)$-connection
 \begin{equation}
  D=d+\bdA.
 \end{equation}
 We will describe this connection by its potential in a trivialisation, $\bdA= A^i \,\gs^i$. 

The associated `twistor space' $\T(M)$ is simply the spinor bundle over $M$, this is an associated vector bundle for our $\SU(2)$-principal bundle: 
\begin{equation}
\C^2 \inj \T(M) \to M.
\end{equation}
We will use adapted local coordinates $\left(x^\mu , \p_{A'} \right)$ to describe this bundle. As always, we will raise and lower spinor indices with the anti-symmetric tensor $\eps_{A'B'}$ (Here this is simply the metric volume form preserved by the $\SU(2)$ action). Having $\SU(2)$ structure group, the $\C^2$ fibers of this bundle come equipped with a hermitian metric that is commonly represented by an anti-linear, anti-involutive map,
 \begin{equation}
  \circonf \colon \left\{\begin{array}{ccc}
\C^2 &\to &\C^2 \\ \p_{A'} &\mapsto &\ph_{A'}
\end{array} \right.
 \end{equation}
such that
\begin{equation}
\ga ,\gb \in \C^2, \qquad \langle \ga , \gb \rangle \coloneqq \gb_{A'}\hat{\ga}^{A'}.
\end{equation}

Making use of the fundamental representation of $\SU(2)$, the $\SU(2)$-connection $D= d + \bdA$ naturally acts as a connection on twistor space :\\
if 
\begin{equation}
 s \left\{ \begin{array}{ccc} M &\to &\T(M) \\ x &\mapsto &\p_{A'}\left(x\right) \end{array} \right.
\end{equation}
is a section of $\T(M)$ then its covariant derivative with respect to $\bdA$ is
\begin{equation}\label{Euclidean Twistor Theory Revisited: Covariant derivativ of section}
	\nabla \p_{A'} = d\p_{A'} - A^{B'}{}_{A'} \;\p_{B'}, \qquad  A^{A'}{}_{B'} \in \su(2).
\end{equation} 
Now we can also re-interpret this last equality in terms of forms: We define the 1-forms $D\p_{A'} \in \gO^1\left(\T(M)\right)$ on the full space of the bundle $\T(M)$ as
\begin{equation}\label{Euclidean Twistor Theory Revisited: Dp definition}
	D\p_{A'} = d\p_{A'} - A^{B'}{}_{A'} \; \p_{B'} \quad \in \gO^1\left(\T(M)\right).
\end{equation}
These are in fact the coordinates of a projection operator, the projection operator on the vertical tangent space to $\T(M)$:
\begin{equation}\label{Euclidean Twistor Theory Revisited: Proj on V of T(M)}
	Proj = D\p_{A'} \otimes \frac{\pa}{\pa \p_{A'}} \quad \in End\left(T \T(M) \right) .
\end{equation}
The kernel of this operator is the \emph{horizontal distribution} associated with the connection $D=d+\bdA$. Thus \eqref{Euclidean Twistor Theory Revisited: Covariant derivativ of section}, \eqref{Euclidean Twistor Theory Revisited: Dp definition} corresponds to the usual dual points of view on connections: either as a differential operators acting on section or as a horizontal distribution on the total space of the bundle.

The associated `projective twistor space' $\PT(M)$ is the projectivised version of $\T(M)$, with Fiber isomorphic to $\CP^1$: 
\begin{equation}
\CP^1 \inj \PT(M) \to M.
\end{equation} 
We will most frequently use homogeneous coordinates $\left(x^{\mu}, \left[\p_{A'} \right] \right)$ to describe this bundle. The main advantage with this notation is that section in $\Oc(n,m)$-bundle over $\CP^1$ (and by extension over $\PT(M)$) are equivalent to functions $f(x,\p_{A'})$ with homogeneity $n$ in $\p_{A'}$ and $m$ in $\ph_{A'}$ .

Similarly k-forms on $\PT(M)$ with values in $\Oc(n,m)$ are uniquely represented by k-forms on $\PT(M)$ with homogeneity $n$ in $\p_{A'}$, $m$ in $\ph_{A'}$ which vanishes on $E = \p_{A'} \frac{\pa}{\pa \p_{A'}}$, $\Eb = \ph_{A'} \frac{\pa}{\pa \ph_{A'}}$: 
\begin{equation}
\begin{array}{ll}
&\ga' \in \gO^k\left(\PT, \Oc(n,m)\right) \\
\Leftrightarrow \quad & \\
&\ga \in \gO^k\left(\T\right) \quad \text{st} \quad E\id \ga = 0, \quad \Eb \id \ga=0, \quad \Ld_{E} \ga = n \ga, \quad \Ld_{\Eb} \ga = m \ga. 
\end{array}
\end{equation}
Here $E$ and $\Eb$ the `Euler vectors', they generate the vertical tangent space of the complex line bundle $\C \inj \T(M) \to \PT(M)$.

As a concrete example,
\begin{equation}
\gt \coloneqq \p_{A'} D \p^{A'} 
\end{equation}
represents a $\Oc(2)$-valued 1-form on $\PT(M)$ but $\ph_{A'} D \p^{A'}$ does not represent a well defined object on $\PT(M)$ as it does not vanish on $Span\left(E,\Eb \right)$.

We can use this fact to define a connection on the $\Oc(n,m)$ bundles.
For suppose $f(x ,\p_{A'})$ represents a section of the $\Oc(n,m)$ bundle, $\Ld_{E}f=n f$, $\Ld_{\Eb} = m f$. Then we can define its covariant derivative as
\begin{equation}
d_{(n,m)} f \coloneqq df + n\;\frac{\ph_{A'} D \p^{A'}}{\pp}\;f - m\;\frac{\p_{A'} D \ph^{A'}}{\pp}\;f
\end{equation}
It is a simple exercise to verify that $E \id d_{(n,m)} f=0$, $\Eb \id d_{(n,m)} f =0$, $\Ld_{E} d_{(n,m)} f = n \;d_{(n,m)} f$ , $\Ld_{\Eb} d_{(n,m)} f = m \; d_{(n,m)} f $ and thus that $d_{(n,m)} f$ indeed represents a $\Oc(n,m)$-valued 1-form on $\PT(M)$.

This connection also preserves the following Hermitian metric on the $\Oc(n,m)$-bundles:
\begin{equation}\label{Euclidean Twistor Theory Revisited: Hermitian metric on O(n)}
	\ga ,\gb \in \Oc(n,m), \qquad \langle \ga , \gb \rangle = \overline{\ga} \;  \gb \; \left(\pp \right)^{-n-m}.
\end{equation}
It is indeed a simple calculation to check that $d_{(n,n)} \left(\pp\right)^{n} =0$. In particular, when restricted to each $\CP^1$ this connection is the natural Chern-connection on $\Oc(n)$ bundle induced by the Kähler structure. 

This connection on $\Oc(n,m)$ bundle over $\PT(M)$ extends to a connection on $\Oc(n,m)$-valued k-forms in the usual way. It is for example instructive to check that,
\begin{equation}\label{Euclidean Twistor Theory Revisited: d tau}
d_{(2)} \gt = F^{A'}{}_{B'}\;\p_{A'}\p^{B'}.
\end{equation}

We thus see that the $\SU(2)$-connection that we started with induces two natural geometric objects on $\PT(M)$: a $\Oc(2)$-valued 1-forms $\gt = \p_{A'}D\p^{A'}$ and a covariant derivative $d_{(n,m)}$ on the $\Oc(n,m)$-bundle over $\PT(M)$.

\subsection{Symplectic and almost Hermitian structure on $\PT(M)$ from a definite connection}

We now restrict ourselves to the case of definite connections \eqref{Definition- Definite Connections and Gravity: Definite Connections}, ie the case where $\tilde{X}^{ij} = F^i\W F^j /_{d^4x}$ is a definite 3x3 conformal metric. This is in fact equivalent to the requirement that no real 3-vector $\left\{v^i\right\}_{i\in 1,2,3}$ is such that $v^i\;F^i$ is a simple 2-form:
\[ 
A\; \text{is a definite connection} \qquad \Leftrightarrow \qquad  \forall v^i \in \R^3 , \; v^i F^i \W v^j F^j = v^i v^j \tilde{X}^{ij} d^4x \neq 0.
\] 

A definite connection on $\PT(M)$ naturally gives a symplectic structure:
\begin{Proposition}{ \emph{Symplectic structure on $\PT(M)$ (Fine and Panov \cite{Fine:2008}) }} \label{Proposition- Euclidean Twistor Theory Revisited: symplectic structure} \mbox{} \\
	If $\bdA$ is a definite connection then $\go_s = \left(n-m\right)^{-1} \left( d_{(n,m)} \right)^2 $ , $n \neq m$, is a symplectic structure on $\PT(M)$.
\end{Proposition}
\begin{proof}\mbox{}\\
	As $d_{n,m}$ is a connection on a line bundle, its curvature 2-form 
 \begin{equation}
	 \go_s = \left(n-m\right)^{-1} \left( d_{(n,m)} \right)^2 = \left(n-m\right)^{-1} d \left(\frac{n\;\ph_{A'} D\p^{A'}-m\;\p_{A'} D\ph^{A'}}{\pp}  \right)
\end{equation}
 is automatically closed. A direct computation shows that,
\begin{equation}\label{Euclidean Twistor Theory Revisited: PT sympleptic structure}
		\go_s = \frac{\p_{A'}D\p^{A'} \W \ph_{B'}D\ph^{B'}}{\left(\pp \right)^2} - F^{A'B'} \frac{\p_{A'} \ph_{B'}}{\pp}
\end{equation} and therefore $\go_s$ is independent of $n$ and $m$. From this last expression one also sees that non degeneracy is equivalent to the definiteness of the connection.
\end{proof}

We also have an almost Hermitian structure obtained by a modification of the classical one described in \cite{Atiyah:1978wi},\cite{Woodhouse85}:

\begin{Proposition}{\emph{Almost Hermitian structure on $\PT(M)$}} \label{Proposition- Euclidean Twistor Theory Revisited: Almost Hermitian structure} \mbox{} \\
	If $\bdA$ is a definite connection then $\PT(M)$ can be given an almost Hermitian structure on $\PT(M)$, ie a compatible triplet $\left(\J_A , \go_A , g_A \right)$ of almost complex structure, 2-form, and a Riemannian metric.  In general this triplet is neither Hermitian ($\J_A$ is not integrable) nor almost Kähler ($\go_A$ is non degenerate but generically not closed). 
\end{Proposition}

\begin{proof}\mbox{}\\
	We first describe how to construct the almost complex structure $\J_A$ on $\PT(M)$ from a definite connection:
	Because the connection is definite, one can make sense of the square root (we take the positive square root) and inverse  of X. Defines $\gS^i = X^{-\frac{1}{2}}{}^{ij} F^j$. By construction $\gS^{i} \W \gS^j \propto \gd^{ij}$. It implies that $\gS_{\pi} = \gS^{A'B'} \p_{A'}\p_{B'}$ is simple, $\gS_{\pi} \W \gS_{\pi}=0$. We now define the almost complex structure by the requirement that $\gO_A = \gt \W \gS_{\pi}$ be a $(3,0)$-form. It makes sense as its kernel,  $\{X\; st\; X\id \gO_A=0\}$, is 3 dimensional and thus can be identified with the $(0,1)$-distribution:
	\begin{equation}\label{Euclidean Twistor Theory Revisited: Hermitian strct: J_A on PT}
		X \in T^{0,1}\PT(M) \quad \Leftrightarrow \quad X\id\; \gt \W \gS_{\pi} =0.
	\end{equation}
	
	This construction has a simple metric interpretation: We already explained how to construct a conformal, non degenerate, Euclidean metric from a definite connection. We will note $e^{AA'}$ the associated null tetrad. It is then easy to see that the construction leading to $\gS^i$ is in fact just an alternative way of constructing $\gS^{A'B'}= \frac{1}{2}e^{A'C}\W e_C{}^{B'}$ ( or equivalently \eqref{Chiral Formulations of GR - Fundations: Sigma def (tetrad)}, see appendix \ref{section: Appdx Spinor conventions} for our conventions on spinors).  In terms of this metric, the $(3,0)$-form reads
	\begin{equation}\label{Euclidean Twistor Theory Revisited: Hermitian strct: gO_A on PT}
	\gO^{3,0}_A = \gt \W \gS^{A'B'}\p_{A'}\p_{B'} = \p_{A'}D\p^{A'} \W e^0{}^{A'}\p_{A'}\W e^1{}^{B'}\p_{B'}.
	\end{equation}
	
	One now comes to the compatible metric on $\PT(M)$. From the definite connection we have a conformal metric. One fixes the scaling freedom by requiring the volume form to be $\frac{3}{2 \gL^2} F^k\W F^k$. We will note $e^{AA'}$ the associated null tetrad.
	This gives a metric on the horizontal tangent space (as defined by $\bdA$), on the other hand the vertical tangent space comes equipped with a metric and altogether this gives the following metric on $\PT(M)$:
	\begin{align}\label{Euclidean Twistor Theory Revisited: Hermitian strct: g_A on PT}
	g_A &= 4R^2\frac{\p_{A'}D\p^{A'} \odot \ph_{B'}D\ph^{B'}}{2\left(\pp \right)^2} + \frac{1}{2}e^{AA'}\odot e_{AA'} \nonumber\\
	&= 4R^2\frac{\p_{A'}D\p^{A'} \odot \ph_{B'}D\ph^{B'}}{2\left(\pp \right)^2} - \frac{e^{AB'}\p_{B'}\odot e_{A}{}^{C'}\ph_{C'}}{\pp}.
	\end{align}
	Here $A \odot B = A \otimes B + B \otimes A$. We leave R, the radius of the fibers, as a parameter but we will see that the Kähler condition will relate it uniquely with $\gL$.
	
	From this, one readily sees that the 2-form,
	\begin{align}\label{Euclidean Twistor Theory Revisited: Hermitian strct: go_A on PT}
	\go_A 
	&= 2iR^2 \left(\frac{\p_{A'}D\p^{A'} \W \ph_{B'}D\ph^{B'}}{\left(\pp \right)^2} -\frac{1}{R^2}\frac{\gS^{B'C'}\p_{B'}\ph_{C'}}{\pp}\right)
	\nonumber\\
	&= 4iR^2\;\frac{\p_{A'}D\p^{A'} \W \ph_{B'}D\ph^{B'}}{2\left(\pp \right)^2} - i\frac{e^{AB'}\p_{B'}\W e_{A}{}^{C'}\ph_{C'}}{\pp} 
	\end{align}
the metric \eqref{Euclidean Twistor Theory Revisited: Hermitian strct: g_A on PT} and the almost complex structure \eqref{Euclidean Twistor Theory Revisited: Hermitian strct: J_A on PT} are compatible.

	This is clear as \eqref{Euclidean Twistor Theory Revisited: Hermitian strct: gO_A on PT},\eqref{Euclidean Twistor Theory Revisited: Hermitian strct: g_A on PT},\eqref{Euclidean Twistor Theory Revisited: Hermitian strct: go_A on PT} are already in the canonical form
\begin{equation}
	\gO_A^{3,0}= dz^1\W dz^2 \W dz^3, \qquad g_A = h_{i\bar{j}}\; dz^i \odot d\bar{z}^{\bar{j}}, \qquad \go_A = i\; h_{i\bar{j}}\; dz^i \W d\bar{z}^{\bar{j}}.
\end{equation}

\end{proof}

Essentially this construction is a variation of the one in \cite{Atiyah:1978wi}. As compared to the classical construction from \cite{Atiyah:1978wi} there are however small differences:\\ 
First the conformal structure is obtained from the connection. \\
Second one does not use the notion of horizontality associated with the (Levi-Civita connection of the) conformal structure but the one given by \emph{our original} $\SU(2)$-connection. In general those two connections differ. The special case where they coincide in fact corresponds to the Einstein case, ie the base metric is Einstein.

For clarity, we expand a little bit on this last point even though this is more related to the pure connection formulation of Einstein equations (that we reviewed in \eqref{Proposition- Definite Connections and Gravity: Pure connection equation}) and somewhat lies out of the main line of development: Suppose that, $\bdA$, the $\SU(2)$-connection that we took as starting point coincide with $\bdA_g$, the (Left chiral part of the) Levi-Civita connection, then their curvature also coincide: $\bdF = \bdF_g$. Now by construction Urbantke metric is such that it makes $\bdF$ self-dual. Therefore $\bdF_g$ is self-dual and this is just the chiral way of stating Einstein equations (cf section \ref{ssection: Chiral Formulations of GR - Fundations}).

In the presence of almost complex structure we can define the projection $\gO^r \to \gO^{(p,q)}$ (here $p+q=r$). For practical purpose, we will write this map as $\ga \mapsto \ga\big|_{(p,q)}$. For example, with the notation of proposition \ref{Proposition- Euclidean Twistor Theory Revisited: Almost Hermitian structure},
\begin{equation}\label{Euclidean Twistor Theory Revisited: Sigma Restrictions}
\gS^{A'B'}\big|_{(2,0)} = \gS\p\p\, \frac{\ph^{A'}\ph^{B'}}{(\pp)^2}, \qquad  \gS^{A'B'}\big|_{(0,2)} = \gS\ph\ph\, \frac{\p^{A'}\p^{B'}}{(\pp)^2},
\end{equation}
\begin{equation*}
\gS^{A'B'}\big|_{(1,1)} = -2\gS\p\ph\, \frac{\p^{(A'}\ph^{B')}}{(\pp)^2},
\end{equation*}
(In order to lighten notations here and thereafter $\gS\pi\pi$ stands for $\gS^{A'B'}\p_{A'}\p_{B'}$ etc).

 It is also natural to introduce Dolbeault operators on the space $\gO^{p,q}\left[n,m\right]$ of $\Oc\left(n,m\right)$-valued $(p,q)$-forms as
\begin{equation}\label{Definition- Euclidean Twistor Theory Revisited: Dolbeault Operators}
\begin{array}{lll}
\pa : \left|
\begin{array}{ccc}
\gO^{p,q}\left[n,m\right] &\to & \gO^{p+1,q}\left[n,m\right] \\ \\
\ga & \mapsto & \left(d_{(n,m)}\ga \right) \big|_{(p+1,q)}
\end{array} \right.
,\quad
\pab : \left|
\begin{array}{ccc}
\gO^{p,q}\left[n,m\right] &\to & \gO^{p,q+1}\left[n,m\right] \\ \\
\ga & \mapsto & \left(d_{(n,m)}\ga \right) \big|_{(p,q+1)}
\end{array} \right.
\end{array}. 
\end{equation}

A this point a natural question is to ask when the almost Hermitian structure introduced in proposition \ref{Proposition- Euclidean Twistor Theory Revisited: Almost Hermitian structure} is Hermitian, i.e $\J_A$ is integrable. 
\begin{Proposition}\label{Proposition- Euclidean Twistor Theory Revisited: Integrability of A} \mbox{}\\
	$\Leftrightarrow$ $\J_A$ is integrable \\
	$\Leftrightarrow$ $\pab \gt =0 $ \\
	$\Leftrightarrow$  $d_{(n)}$ is compatible with $\J_A$, ie $\left(d_{(n)}\right)^2 \big|_{(0,2)} =0$ \\
	$\Leftrightarrow$ $\gt \W d\gt \W d\gt=0$ \\
	$\Leftrightarrow$ $A^i$ is perfect : $F^i \W F^j \propto \gd^{ij} d^4x$.\\
	
	It follows that under this conditions the $\Oc(n)$-bundles are holomorphic with Hermitian metric \eqref{Euclidean Twistor Theory Revisited: Hermitian metric on O(n)} and $d_{(n)}$ is the associated Chern connection.
\end{Proposition}
We recall that by proposition \eqref{Proposition- Definite Connections and Gravity: Perfect connection} a perfect connections is the connection of an anti-self-dual Einstein metric. In particular, under the assumption of proposition \ref{Proposition- Euclidean Twistor Theory Revisited: Integrability of A}: \[ \text{\emph{The Urbantke metric with volume form $\frac{3}{2 \gL^2} F^k \W F^k$ is anti-self-dual Einstein}}. \]

\begin{proof}\mbox{}\\
	
	We now prove that each point taken separately is equivalent to perfectness of the connection.
	
	It is easy to check that $\gt \W d\gt \W d\gt = \gt \W F^{A'B'} \W F^{C'D'} \p_{A'}\p_{B'}\p_{C'}\p_{D'}$ and thus $\gt \W d\gt \W d\gt=0$ is directly equivalent to the perfectness of the connection.
	
	In section \ref{ssection: Definite Connections and Gravity}, we saw that $F^i = s \, \sqrt{X}^{ij} \gS^j$. Thus we can write
\begin{equation}\label{Euclidean Twistor Theory Revisited: Proof: F decomposition}
	 F^{A'B'} = \Psi^{A'B'}{}_{C'D'} \gS^{C'D'} + \gl(x)\gS^{A'B'}\quad \text{with} \quad \Psi^{A'B'C'D'} = \Psi^{(A'B'C'D')}.
\end{equation}
	It was also explained in section \ref{ssection: Definite Connections and Gravity} that the self-dual Einstein equations, i.e perfectness of the connection, are equivalent to $\Psi=0$. Then our choice of volume form $\nu = 2\gS^i \W \gS^i = \frac{3}{2 \gL^2} F^k \W F^k$ gives $ \gl(x) = s |\gL|$.
	
	Taking \eqref{Euclidean Twistor Theory Revisited: d tau} and \eqref{Euclidean Twistor Theory Revisited: PT sympleptic structure} together with \eqref{Euclidean Twistor Theory Revisited: Sigma Restrictions}, \eqref{Definition- Euclidean Twistor Theory Revisited: Dolbeault Operators} and \eqref{Euclidean Twistor Theory Revisited: Proof: F decomposition} one easily shows that 
	\[ \left(d_{(2)}\gt \right)\big|_{0,2} = \Psi\p\p\p\p \; \frac{\gS \ph\ph}{(\pp)^2},\qquad \pab \gt = \Psi\p\p\p\ph \; \frac{\gS \p \ph}{(\pp)^2},\]
	and
	\[
	\frac{1}{n}\left(d_{(n)}\right)^2 \big|_{(0,2)}= \Psi\p\p\p\ph \frac{\gS \ph\ph}{(\pp)^2}.\]
	Therefore $d_{(2)}\gt \big|_{(0,2)} =0$,  $\pab \gt = 0$ and $\left(d_{(n)}\right)^2 \big|_{(0,2)}=0$  are separately equivalent to $\Psi =0$, i.e to the perfectness of the connection.
	
	Finally, all is left to show is that integrability of $\J_A$ is equivalent to the perfectness of the connection. However integrability is equivalent to having both $d\gt \big|_{(0,2)} =0$ and $d\left(e^{AA'}\p_{A'}\right)\big|_{(0,2)}$ and thus imply perfectness of the connection. On the other hand, if the connection is perfect then 
	\begin{equation*}
	d\left(e^{BB'}\p_{B'}\right)\big|_{(0,2)} = \left(d_A e^{BB'} \right)\big|_{(0,2)} \p_{B'} + e^{BB'} \W D\p_{B'}\big|_{(0,2)} = \left(d_A e^{BB'} \right)\big|_{(0,2)} \p_{B'}
	\end{equation*}
	 holds identically as a result of $\bdA$ being the self-dual connection associated with the tetrad.
\end{proof}

As already emphasised in the introduction, the main difference with the traditional results from \cite{Atiyah:1978wi} is that integrability is not only related to the anti-self-duality but is irremediably linked to Einstein equations. This is because in the construction described in \cite{Atiyah:1978wi} one is only interested in a conformal class of metric while here the use of the connection automatically fixes the `right scaling' that gives Einstein equations.

It is also natural to ask under which condition the almost hermitian structure is almost Kähler and Kähler. In fact those two situations necessarily come together but depends on the sign of the connection: 

\begin{Proposition} \label{Proposition- Euclidean Twistor Theory Revisited: Kahler condition}\mbox{}\\
	$g_A$ is Kähler \\
	$\Leftrightarrow$ $\J_A$ is integrable, $s=1$ and $R^2 = \frac{3}{|\gL|}$\\
	$\Leftrightarrow$ $\go_A$ is closed (an thus sympleptic) \\
	$\Leftrightarrow$ $\go_A = 2i R^2 \;\go_s$ 
\end{Proposition}
\begin{proof} \mbox{}\\
	$\go_A = 2i R^2 \;\go_s$ is easily shown to be equivalent to $F^i = \frac{1}{R^2} \gS^i$. This is only possible if the connection is perfect with positive sign. The same is true for the closeness of $\go_A$, ie direct computations shows the equivalence of the last two point with $F^i = \frac{1}{R^2} \gS^i$.
	
	Now, from proposition \ref{Proposition- Euclidean Twistor Theory Revisited: Integrability of A} perfectness of the connection is equivalent to integrability. Incidentally one sees from $F^i = \frac{1}{R^2} \gS^i$ that the metric associated with the connection is self-dual Einstein with cosmological constant $\gL = \frac{3}{R^2}$.
\end{proof} 

When the connection is perfect, the Hermitian structure that we described restrict to the usual Hermitian structure on twistor space constructed from an Instanton. The discussion on the sign of the connection parallel the well known fact that this Hermitian structure can be made Kähler only if the cosmological constant is positive.

\subsection{The Mason-Wolf action for self-dual gravity}\label{ssection: The Mason-Wolf action for self-dual gravity}

In \cite{Mason&Wolf09}, L.Mason and M.Wolf described a twistor action for self-dual gravity. It is an action for an $\Oc(2)$-valued 1-form $\gt$ and a $\Oc(-6)$-valued 1-form $b$ on some 6d real manifold, the  `projective twistor space'. It essentially used a new version of the non linear graviton theorem relying on the equation $\gt\W d\gt\W d\gt=0$. This equation was understood as a sufficient condition for the integrability of a certain almost complex structure and thus, relying on Penrose-Ward Non-Linear-Graviton theorem \cite{Penrose:1976js} \cite{Ward:1980am}, as describing some Einstein anti-self-dual space-time. The Mason-Wolf action implements this constraint with a Lagrange multiplier:
\begin{equation}\label{Euclidean Twistor Theory Revisited: Action: Mason-Wolf}
	S\left[\gt, b\right] = \int_{\PT} b\W \gt\W d\gt\W d\gt 
\end{equation}
Even thought the logic that lead to this Lagrangian was somehow different, in retrospect one sees that this Lagrangian could have been guessed from the description of self-dual gravity in terms of perfect connections. Indeed, as already explained in section 2, in terms of $\SU(2)$-connections the equations for self-dual gravity read $F^{(A'B'} \W F^{C'D')}=0$ and therefore we can easily obtain an action for self-dual gravity by implementing this constraint by a Lagrange multiplier:
\begin{equation}\label{Euclidean Twistor Theory Revisited: Action: Space-time SD gravity FF}
	S\left[B, A\right] = \int_{M} B_{A'B'C'D'} F^{A'B'}\W F^{C'D'},
\end{equation}
where the $B$ field is completely symmetric, $B^{A'B'C'D'} = B^{(A'B'C'D')}$.

Now, as discussed before the natural `Penrose transform' of a $\SU(2)$-connection is the $\Oc(2)$-valued 1-form on $\PT(M)$,
  \begin{equation}
\gt = \p_{A'}\left(d\p^{A'} + A^{A'}{}_{B'} \p^{B'}\right)
\end{equation}
 We also take the Penrose transform of $B$ to be
  \begin{equation}
  B^{A'B'C'D'} = \int_{\CP^1} \p^{A'}\p^{B'}\p^{C'}\p^{D'}\; b \W \gt  
 \end{equation}
  with $b$ a $\Oc(-6)$ valued (0,1)-form on $\PT(M)$. This is just the usual Penrose transform for massless fields, see eg \cite{Woodhouse85}.  We recall from the previous discussion that \begin{equation}
  \gt \W d\gt \W d\gt = \gt \W F^{A'B'}\p^{A'}\p^{B'}\W F^{C'D'}\p^{C'}\p^{D'}.
  \end{equation}

From this one readily sees that the Mason-Wolf action \eqref{Euclidean Twistor Theory Revisited: Action: Mason-Wolf} is the immediate generalisation of \eqref{Euclidean Twistor Theory Revisited: Action: Space-time SD gravity FF}. 

The aim of the sub-section \ref{ssection: the NLGT revisited} is to make the relation between the Mason-Wolf action and the connection description of Einstein self-dual connections even more precise by giving a new proof of the non linear graviton theorem that emphasise this relation.

Our proof will indeed make it clear that this version of the non linear graviton theorem has a strong `connection' flavour. It might therefore suggest new types of generalisation to full gravity. We discuss some of those in section \ref{section: Discussion on the would be `Twistor action for Einstein gravity'} where we will described strategies towards a twistor action for full gravity. 

\subsection{The Non linear graviton theorem revisited} \label{ssection: the NLGT revisited}

Up to now we constructed different geometrical structure on $\PT(M)$ from a definite connection. In particular we saw that $\PT(M)$ can be given a Kähler structure when $\gt \W d\gt \W d\gt =0$, with $\gt = \p_{A'} \left( d\p^{A'} + A^{A'}{}_{B'}\p^{B'}\right)$. 

We are now interested in the reverse problem: We take `projective twistor space' $\PTc$ to be an oriented manifold diffeomorphic to $\R^4\times S^2$ together with a 1-form $\gt \in \gO^1_{\C}\otimes L $ with values in a line bundle $L$ over $\PTc$. We suppose this line bundle to be such that its restriction to each $S^2$ has Chern class 2. This is enough to define an almost complex structure $\J_{\gt}$ on $\PTc$ as we now describe.

 We thank L.Mason for important discussions and suggestions that greatly contributed to this presentation.

\paragraph{The almost complex structure $\J_{\gt}$}\mbox{}\\
We first introduce the 4-dimensional  `horizontal distribution' $ H\subset T_{\R}\PTc $ defined as the kernel of $\gt$, $H = Ker\left(\gt\right) $. 

We then determine $\gl\in C^{\infty}(\PTc)$ as
\begin{equation}\label{Euclidean Twistor Theory Revisited: lambda def}
	\gt \W \gtb \W \left( \gl d\gt + d\gtb \right)^2 =0.
\end{equation}
This is a quadratic equation for $\gl$. We then construct $a$, a connection on the $L$ bundle, defined modulo the addition of multiple of $\gt$ and $\gtb$ by requiring, 
\begin{equation}\label{Euclidean Twistor Theory Revisited: a def}
	\gtb \W \left( d\gtb + \gl \left(d\gt + a \W \gt  \right)\right)^2 =0.
\end{equation}
This is in fact linear in $a$ and has the right number of components to determine $a$ modulo $\gt$ and $\gtb$.
From all this we define the complex 3-form,
\begin{equation}\label{Euclidean Twistor Theory Revisited: J from tau def}
	\gOb = \gtb \W \left( d\gtb + \gl\,d_a\gt \right).
\end{equation}
This 3-form in turn defines an almost complex structure, $\J_{\gt}$: We just define the holomorphic tangent space to be the kernel of $\gOb$,
\begin{equation}
X \in T^{1,0}\PTc \quad \overset{def}{\Longleftrightarrow} \quad X \id \gOb =0.
\end{equation}
From \eqref{Euclidean Twistor Theory Revisited: a def} we see that the Kernel of $\gO$ indeed is 3-dimensional as required for an almost complex structure. Note that this definition of $\J_{\gt}$ is equivalent to requiring that $\gOb$ is $(0,3)$. In particular its complex conjugate $\gO$ is (3,0).

\paragraph{Spacetime from $\J_{\gt}$} \mbox{}\\
Having constructed an almost complex structure, $\J_{\gt}$ on $\PTc$ we are now in a similar situation as in \cite{Mason05} where the almost complex structure is taken as a starting point.

 Following the same steps as in this reference \emph{we can construct a Euclidean `space-time' $M$ from $\left(\PTc, \J_{\gt}\right)$}. Then $\PTc$ has the structure of a fiber bundle over $M$ : $\CP^{1} \inj \PTc \to M$. Twistor space $\mathcal{T}$ is taken as the total space of a special line bundle over $\PTc$. We here recall how this works for completeness.\\

We first introduce a conjugation $\circonf \from \PTc \to \PTc $, $\circonf^2 = 1$, that reverses $\J_{\gt}$, i.e $\circonf^* \J_{\gt} = - \J_{\gt}$. We also assume that this conjugation has no fixed points. This is a common in twistor theory and will lead to a Euclidean Space-time, the other alternative (existence of fixed points for $\circonf$) would lead to Lorentzian signature.

We now take as \emph{`complexified space-time'} $\Mc$ the moduli space of pseudo-holomorphic rational curves in $\PTc$, ie the space of embedded $S^2$ in $\PTc$ in the same topological class as the $S^2$ factors in $\PTc \simeq \R^4 \times S^2$ such that $\J_{\gt}$ leaves the tangent space invariant and thus inducing a complex structure on these embedded 2-spheres. Theorems in McDuff and Salamon \cite{McDuff&Salamon2004} imply that $\Mc$ exists and is 8-dimensional if $\J_{\gt}$ is close to the standard complex structure on a neighbourhood of a line in $\CP^3$. We assume this condition to be satisfied. This can be done by requiring that our 1-form $\gt$ is close to the standard holomorphic 1-form with values in $\Oc(2)$ on $\CP^3$.

The conjugation $\circonf$ induces a conjugation on $\Mc$, $\circonf \from \PTc \to \PTc$ and we define our \emph{Euclidean space-time} $M$ as the 4-dimensional fixed point set of $\circonf$ on $\Mc$. There is then a natural projection $P \from \PTc \to M$ as a consequence of the fact that from our assumption that there will be a unique rational curves in $\PTc$ through $Z$ and $\hat{Z}$. By construction our projective twistor space $\PTc$ now is the total space of a fibre bundle over $M$ with fibre $\CP^1$: $\CP^1 \inj \PTc \xto{P} M$.

We will also assume that $\J_{\gt}$ is such that the canonical bundle $\gO^{3,0}$ has Chern class $-4$ on each $S^2$ in $\PTc$. This will be the case if we construct $\gt$ by a small deformation of the standard holomorphic 1-form with values in $\Oc(2)$ on $\CP^3$.

We then define the associated twistor space $\Tc$ to be the fourth root of the canonical bundle. It is thus a complex line bundle over $\PTc$, $\C \inj \Tc \xto{\Pi} \PTc$. We denote the complex line bundle $\left(\gO^{3,0}\right)^{-\frac{n}{4}}$ by $\Oc(n)$. When restricted to each $\CP^1$ fibres in $\PTc$, these bundles will restrict to the usual $\Oc(n)$ holomorphic bundle on $\CP^1$ and thus the notation is coherent. We can now think of $\Tc$ as a complex rank two vector bundle over $M$ with structure group $\SU(2)$, $\C^2 \inj \Tc \xto{P'} M$.

\paragraph{A non linear graviton theorem} \mbox{}\\
We now give a new proof of the (euclidean) non-linear-graviton theorem. As explained in introduction, the essential result of this theorem already appeared in \cite{Mason&Wolf09} but the presentation that we make here is original.\\

Introduce coordinates that form a trivialisation of $\mathcal{T}$, $\{x^{\mu}, \p_{A'} \}$.  $\p_{A'} \; \text{with} \; \text{\footnotesize A'} \in\{0,1\}$ are linear coordinates on the fibres of $\C^2 \inj \Tc \xto{P'} M$ and $x^{\mu}$ are local space-time coordinates on the base.\\ Then,
\begin{Proposition}\label{Proposition- Euclidean Twistor Theory Revisited: Non Linear graviton} \mbox{}\\
	(i)  \quad 	$\J_{\gt}$ is integrable\\
	(ii) \quad $\Leftrightarrow$ $\gt \W d\gt \W d\gt =0$ \\
	(iii)\quad $\Leftrightarrow$ $\gt = \gc \left(\p_{A'}d\p^{A'} + A(x){}^{A'}{}_{B'}\p^{B'}\p_{A'} \right)$ with $\gc \in \C^{\infty}\left(\PTc \right)$ and $A^{A'}{}_{B'}$ a perfect connection on $M$.
\end{Proposition}

\begin{proof} \mbox{} \\
	We now prove $(i) \Rightarrow (ii) \Rightarrow (iii) \Rightarrow (i)$.\\
	$(i) \Rightarrow (ii)$\\
	By construction, $\gt \W \left(d\gt + \bar{\gl} d_{\bar{a}} \gtb \right)^2 = \gO \W \left( d\gt + \bar{\gl} d_{\bar{a}} \gtb \right)=0$, integrability means that $\gO \W d\gt =0$ and thus $\bar{\gl} \;  \gO \W d_{\bar{a}}\gtb = 0$. It follows that either $\gl=0$ or $d_{\bar{a}}\gtb \big|_{(0,2)} =0$.
	If $\gl=0$ then $\gt\W d\gt \W d\gt =0$. Suppose $d_{\bar{a}}\gtb \big|_{(0,2)} =0$, integrability implies that $ d_{\bar{a}} \gtb \big|_{(2,0)} = 0$ and therefore $d_{\bar{a}} \gtb \in \gO^{1,1}$. It follows that both $d_{\bar{a}} \gtb \in \gO^{1,1}$ and $d_{a} \gt \in \gO^{1,1}$. However this is in contradiction with $\gt \W \left( d_{a}\gt + \gl d_{\bar{a}} \gtb \right) \in \gO^{3,0}$. \\
	$(ii) \Rightarrow (iii)$\\
	If $\gt \W d\gt \W d\gt =0$ then by construction $\gt \W d\gt \in \gO^{3,0}$. We now take $\gz$ to be coordinates on $\CP^1$, $\pa_{\gzb}$ is the anti-holomorphic vertical tangent vector. It follows that $\pa_{\gzb} \id d\gt \propto \gt$. Using coordinates adapted to the trivialisation we can write $\gt =\gl\left( d\gz + A_{\mu} dx^{\mu}\right)$ and $\gt\W \gzb \id d\gt =0$ implies $\pa_{\gzb}A_{\mu}=0$. $A_{\mu}$ being $\Oc(2)$ valued this last relations implies, by a generalisation of Liouville's theorem, $A = A(x){}^{A'}{}_{B'}\p^{B'}\p_{A'}$. Now  \ref{Proposition- Euclidean Twistor Theory Revisited: Integrability of A} implies that $A^{A'}{}_{B'}$ is perfect.\\
	$(iii) \Rightarrow (i)$\\
	Starting with a perfect connection $\bdA$, we have from Proposition \ref{Proposition- Euclidean Twistor Theory Revisited: Integrability of A} that $\gt \W d\gt \W d\gt =0$. From the definition of the almost complex structure, this implies $\gt \W d\gt = \gt \W F\p\p \in \gO^{(3,0) }$. For a perfect connection, $F^i = s\, \frac{\gL}{3} \gS^i $, and the almost complex structure is therefore the same as in proposition \ref{Proposition- Euclidean Twistor Theory Revisited: Almost Hermitian structure}. Finally, by proposition \ref{Proposition- Euclidean Twistor Theory Revisited: Integrability of A} perfectness of the connection implies integrability.
\end{proof}

From this point of view, the `non-linear graviton theorem', with central equation $\gt\W d\gt\W d\gt=0$, can therefore be understood as a deep generalisation of the description of self-dual gravity in terms of perfect connections $F^i\W F^j = \frac{\gd^{ij}}{3} F^k\W F^k$, cf \eqref{Proposition- Definite Connections and Gravity: Perfect connection}. As we already reviewed, full gravity can be described in terms of $\SU(2)$-connection only (cf proposition \eqref{Proposition- Definite Connections and Gravity: Pure connection equation}) and this is therefore suggestive of a twistor description of full gravity in terms of the 1-form $\gt$ only. 

\section{Discussion on the would be `Twistor action for Einstein gravity'} \label{section: Discussion on the would be `Twistor action for Einstein gravity'}

In \cite{Mason05} two new variational principles for Yang-Mills theory and conformal gravity based on fields living on twistor space were presented. The fact that the fields which appear in this action live on a 6d manifold (`projective twistor space') is compensated by new symmetries of the action and the propagating degrees of freedom thus remain the same as in the usual Yang-Mills or Conformal gravity action as expected. 
One of the nice features of these actions is that they give a natural explanation for why there is a MHV formalism for Yang-Mills and Conformal gravity. Because of the extra symmetries that these action enjoys (as compared to the space-time action) they allow to choose a special gauge ( referred to as CSW, for Cachazo-Svrcek-Witten, gauge) that makes a MHV formalism manifest (cf \cite{Jiang08}, \cite{Adamo:2013cra}, \cite{Adamo:2011pv} and \cite{Adamo:2013tja}).
If such an action existed for GR one could expect the same phenomenon and it could serve as a proof for the existence (or the obstruction to the existence) of a MHV formalism for GR.

As already described in section \ref{ssection: The Mason-Wolf action for self-dual gravity} a twistor action for self-dual gravity was presented in \cite{Mason&Wolf09}. Then, in \cite{Adamo:2013tja} a conjectured twistor action for full gravity was proposed. However, in spite of the many interesting features of this conjectured twistor action, some geometrical understanding is lacking, mainly because it is formulated around a fixed background, and this makes it unclear whether it actually describe GR or not.

Both the twistor action for Yang-Mills and Conformal gravity where obtained by a generalisation of the respective space-time action. We very briefly sketch how this works for Yang-Mills, but refer the reader to \cite{Mason05} for details on the construction. This is of interest for us because we will see that, together with the description of metric in terms of connection described in section \ref{section: Chiral formulation of gravity} it has an immediate generalisation to gravity.

\subsection{The Twistor action for Yang-Mills from the Chalmer-Siegel action}\label{ssection: The Twistor action for Yang-Mills from the Chalmer-Siegel action}

In \cite{Mason05}, the Chalmer-Siegel \cite{Chalmers:1996rq} action for Yang-Mills was taken as a starting point on the way to a twistor action:
\begin{equation}
	S\left[A, B\right] = \int_M Tr\left( B \W F - \frac{\eps}{2} B \W B \right)
\end{equation}
where $B$ is taken to be a lie algebra valued \emph{self-dual} 2-form, ie $B = B_{A'B'} \gS^{A'B'}$. Here $\gS^{A'B'}$ is a basis of self-dual 2-forms associated with a fixed background flat metric and constructed as in \eqref{Chiral Formulations of GR - Fundations: Sigma def (tetrad)}. As already described, the Euclidean twistor space is the total space of the primed spinor bundle over $M$, the almost complex structure on $\PT$ is given by taking the $(3,0)$-form on $\PT$ to be $\gO = \gt \W \gS^{A'B'}\p_{A'}\p_{B'} = \p_{C'}d\p^{C'}\W \gS^{A'B'}\p_{A'}\p_{B'}$. 

An interesting feature of this action is that for $\eps=0$ we are left with an action for self-dual Yang-Mills. This is a key point to make contact with twistor theory as self-dual Yang-Mills solutions are fully understood in terms of geometry of the twistor space through the Ward transform \cite{Ward:1977ta}.

If we take the Penrose transform of $B_{A'B'}$ to be 
\begin{equation}
B_{A'B'} = \int_{\CP^1} \p_{A'}\p_{B'} \;b\W \gt
\end{equation}
(where $b$ is a $(0,1)$-form on $\PT$ with values in $\Oc(-4)$) and plug it into the action, we see that it is suggestive of the twistor action for Yang-Mills \cite{Mason05}:
\begin{equation}
	S\left[a, b\right] = \int_{\PT} Tr\left( b \W f \W \gO \right) - \frac{\eps}{2} \int_{M\times \CP^1 \times \CP^1} Tr\left(b_1 \W \gt_1 \W b_2 \W \gt_2 \left(\p_{1}{}_{A'} \p_{2}{}^{A'}\right)^2 \right).
\end{equation}
Where now $a$ is taken to be a (0,1) $\SU(N)$-connection of a Yang-Mills bundle \emph{over $\PT$} and $f \in \gO^{0,2}\left(\PT \right)$ its curvature. For $\eps=0$ varying the action with respect to $b$ gives $f=0$ and thus gives this bundle the structure of a holomorphic bundle over $\PT$. By a theorem from Ward \cite{Ward:1977ta}, this is equivalent to self-dual Yang-Mills equations see also \cite{Woodhouse85} for Euclidean methods in twistor theory. What's more, it turns out that this action describes full Yang-Mills for $\eps\neq 0$. Again, the aim here is just to sketch how this action is constructed and refer to \cite{Mason05} for a proper discussion.

\subsection{A first twistor ansatz... and why it fails}

We here would like to take as starting point the following space-time action:
\begin{equation}
	S[A, \Psi] = \int \Psi^{ij}\; F^i\W F^j + \sum_{k=0}^{\infty} \left(\frac{3}{\gL}\right)^{k+1} \; \left(\Psi^{k+2}\right)^{ij} F^i \W F^j
\end{equation}
This action is in fact an action for gravity. We explain in Appendix \ref{section: Appdx A Review of Chiral Lagrangians for Gravity} how it can be derived by integrating fields from the Plebanski action.

This action, which does not seem to have attracted much attention up to now, has the following interesting interpretation: in the limit where $\gL$ goes to infinity we recover an action for anti-self-dual gravity. For a finite $\gL$ however this action describe full GR as an interacting theory around the anti-self-dual background with the cosmological constant playing the role of coupling constant. This parallels the Chalmers-Siegel action for Yang-Mills.

It suggests the following twistor ansatz,

\begin{align}
	S\left[\gt, \psi \right] &= \int_{\PTc} \psi \W \gt \W d\gt \W d\gt \\
	& \hspace{1cm}  + \sum_{k=0}^{\infty} \left(\frac{3}{\gL}\right)^{k+1}\;\smashoperator{\int_{\hspace{1.75cm} M \times \CP^1 \times \CP^1}} \psi_1 \W \gt_1 \W d\gt_1 \W \psi_2 \W \gt_2 \W d\gt_2  \left(\Psi\right)^k{}^{A'B'}{}_{C'D'}\;\p_1{}^{C'}\p_1{}^{D'}\p_2{}_{A'}\p_2{}_{B'} \nonumber
\end{align} 
where  $\gt$ is a $\Oc(2)$-valued 1-form on $\PTc$. As we already described in \ref{ssection: the NLGT revisited} such a 1-form is enough to construct an almost complex structure $\J_{\gt}$ on $\PTc$ and to give it a fibre bundle structure over some space-time $M$, $\CP^1 \inj \PTc \to M$.

This action also contains $\psi\in \gO^1_{\C} \otimes \Oc\left(-6\right)$ a 1-form on $\PTc$ with values in $\Oc(-6)$, its Penrose transform is $\Psi(x){}^{A'B'}{}_{C'D'}$:
\begin{equation}
\Psi\left(x\right){}^{A'B'}{}_{C'D'} = \int_{\CP^1_x} \psi\W \gt \;\p{}^{A'}\p{}^{B'}\p{}_{C'}\p{}_{D'}
\end{equation}
where $\CP^1_x = P^{-1}(x)$ is the fibre above $x\in M$.

Interestingly, when $\gL$ goes to infinity one recovers the Mason-Wolf action for self-dual gravity described in section \ref{ssection: The Mason-Wolf action for self-dual gravity}. What's more, truncating the infinite sum to the first term one recovers an action that looks like a background independent version of the twistor action conjecture in \cite{Adamo:2013tja}.

Unfortunately, despite those encouraging features, one cannot prove that this twistor action is related to gravity. To do so one would hope that varying this action with respect to $\psi$ would give enough field equations to recover a $\SU(2)$-connection from $\gt$ as was the case in our proof of the non-linear graviton theorem  \ref{ssection: the NLGT revisited}. However this does not seems to be the case here. We are thus unable to make contact with a space-time counter part of this action and the interpretation of the fields equations remains obscure.

\subsection{A new action for Gravity as a background invariant generalisation of the Chalmers-Siegel action.}\label{ssection: A new action for Gravity as a background invariant generalisation of the Chalmers-Siegel action.}

Let's now come back to the Chalmer-Siegel action and consider the special case of a $\SU(2)$-connection:
\begin{equation}\label{Discussion on the would be `Twistor action for Einstein gravity': Action: Chalmer-Siegel}
	S\left[A, B\right] = \int_M B^i_{A'B'} \gS^{A'B'}\W F^i - \frac{\eps}{2} B^i_{A'B'}B^i{}_{C'D'} \gS^{A'B'}\W \gS^{C'D'}
\end{equation}
Here $\gS^{A'B'}$ is a basis of self-dual 2-forms associated with a fixed background flat metric and constructed as in \eqref{Chiral Formulations of GR - Fundations: Sigma def (tetrad)}. This action has obviously nothing to do with an action for gravity as a background metric is present.

However, as explained in section \ref{section: Chiral formulation of gravity}, a definite $\SU(2)$-connection is enough to define a conformal class of metric. If we now choose a representative in the conformal class, and use it to parametrise the $\gS$'s, $\gS_A = \gS(g_A) = \gS(A) $, the action \eqref{Discussion on the would be `Twistor action for Einstein gravity': Action: Chalmer-Siegel} becomes background independent:
\begin{equation}
S[A^i, B^{ij}] = \int B^{ij} \gS_A^i\W F^j  - \frac{\eps}{2} B^{ij}B_{ik}\; \gS_A^j\W \gS_A^k
\end{equation}
If we take $B^{ij}$ to be unconstrained, the action ends up to be topological. However if we take $B^{ij}$ to be traceless with the good choice of volume form then the action happens to describe gravity in the pure connection formulation:
\begin{Proposition}\label{Proposition- Discussion on the would be `Twistor action for Einstein gravity': Action: generalised CS}\mbox{}\\
	The action
\begin{equation}	 
	S[A^i, B^{ij}] = \int B^{ij} \gS_A^i\W F^j  - \frac{\eps}{2} B^{ij}B_{ik}\; \gS_A^j\W \gS_A^k
\end{equation}
	,with $\gS_A^i$ the basis of orthonormal self-dual 2-form associated with Urbantke metric with volume $\frac{1}{2\gL^2}\left(tr \sqrt{F\W F}\right)^2$ and $B^{ij}$ a traceless matrix,  describes the vacuum solution of Einstein equations with non zero cosmological constant. What's more for $\eps=0$ this action describes anti-self-dual gravity.
\end{Proposition}
\begin{proof}
	
	By construction the $\gS_A$'s are such that,
\begin{equation}
	F^i = M^{ij} \gS_A^j.
\end{equation}
	Our choice of volume form,
\begin{equation}
	\frac{1}{3}\gS^i_A \W \gS^i_A = \frac{1}{\gL^2}\left(tr \sqrt{F\W F}\right)^2,
\end{equation}
	is such that $Tr M = \gL$ is a constant.
	
	Now, varying the action with respect to $B$, we get 
\begin{equation}
	 M^{ij}\big|_{trace-free} = \eps B^{ij} 
\end{equation}
	which is equivalent to 
\begin{equation}
	F^i =  \left(\eps B^{ij}+ \frac{\gL}{3} \gd^{ij}\right)\gS^j.
\end{equation}
	For $\eps=0$, these are the equations for self-dual gravity in terms of connection.
	For $\eps\neq 0$ we can solve for $B$, plugging this back into the action we obtain
\begin{equation}
	S\left[A\right] = \frac{1}{\eps} \int \frac{1}{2} F^i\W F^i - \frac{1}{6} \left(tr \sqrt{F\W F}\right)^2.
\end{equation}
	Up to a topological term, this is just the pure connection action for gravity \cite{Krasnov:2011pp}, see also appendix \ref{section: Appdx A Review of Chiral Lagrangians for Gravity}.
\end{proof}

\subsection{Discussion on a second ansatz}

The action in proposition \eqref{Proposition- Discussion on the would be `Twistor action for Einstein gravity': Action: generalised CS} looks like a promising starting point to construct ansatz for twistor action for gravity. It indeed has many appealing features. First it explicitly separates the self-dual sector ($\eps=0$) of the theory from the full theory ($\eps \neq 0$). Second it superficially looks like the space-time counterpart of the twistor action conjectured in \cite{Adamo:2013tja}. Finally, as explained in the previous section the $\SU(2)$-connection on $M$ can naturally be lifted as a 1-form $\gt$ on $\PTc$. Starting with an action of this type would again allow to use the machinery described in section \ref{section: Euclidean Twistor Theory Revisited}.

However as to now, despite many attempts from the author of this paper, none of the ansatz that are suggested by this action seem to lead to an interesting gravity action. We describe here one attempt that seemed at some point the most promising to the author. We will see that it can indeed eventually lead to a certain variational principle in twistor space but at the expense both of technical complications and the addition of an unnatural constraint. Thus the result seems both too complicated to be directly useful (let say for computing scattering amplitudes) and to anaesthetic to be otherwise appealing. However, on the way the interested reader should get some glimpses on the type of difficulties that one faces when one tries to construct such a variational principle in twistor space.

The essential idea here is it that we now would like to construct an action of the type $S\left[\gt, \psi \right]$ on $\PTc$, some real 6d manifold. First using the results from section \ref{section: Euclidean Twistor Theory Revisited}, one can construct an almost complex structure $\J_{\gt}$ which in turn allows to construct some space time $M$, giving $\PTc$ a fibre bundle structure $\CP^1 \inj \PTc \xto{\pi} M $. A look at the action \ref{Proposition- Discussion on the would be `Twistor action for Einstein gravity': Action: generalised CS} then suggests the following `twistor ansatz':
\begin{align}\label{Discussion on the would be `Twistor action for Einstein gravity': twistor ansatz 2}
	S\left[\gt, \psi \right]&= \int_{\PT} \psi \W \gt \W d\gt \W \gS_{\gt} + \frac{\eps}{2} B^{A'B'} \W B_{A'B'}
\end{align}
where $\gS_{\gt}$ should be constructed from $\gt$ only,
\begin{equation}
\psi \in \gO^{1}_{\C}\otimes \Oc(-6)
\end{equation}
and 
 \begin{equation}
B(x)^{A'B'}= \int_{\CP_X^1} \p^{A'}\p^{B'} \; \psi \W \gt \W \gS_{\gt}.
\end{equation}

Where in this last line one should integrate over $\pi^{-1}(x) \simeq \CP^1$.

An appealing feature of actions of this type is that, linearising around a given background (let say describing flat space-time) we obtain $\gd\psi \in H^{0,1}\left(\PTc, \Oc(-6)\right)$ and $\gd \gt \in H^{0,1}\left(\PTc, \Oc(2)\right)$ which are then naturally interpreted as the Penrose transform of a propagating self-dual $\Psi_{A'B'C'D'}$ and anti-self-dual $\Psi_{ABCD}$ gravitons.

The difficult part now is to make sense of $\gS_{\gt}$. We propose the following.
In section \ref{section: Euclidean Twistor Theory Revisited} we partly defined a connection $a$ on $\Oc(2)$,  through the relation
\begin{equation}
	\gtb \W \left( d\gtb + d_a \gt\right)^2 =0
\end{equation}
however as for now it is only defined up to multiple of $\gt$, $\gtb$. If we require as some non-degeneracy condition that $\gt$ does not vanish on the $\CP^1$ that fibers $\PTc$, we can then completely fix $a$ by requiring that the $\Oc(2)$-connection that is induces on each $\CP^1$ fibres is the Levi-Civita connection of the Kahler metric on $\CP^1$. Now that $a$ is completely defined we have access to its curvature $\left( d_a\right)^2$.

Consider the following triple of 2-forms: $\left(d_a\gt,d_a\gtb, \left( d_a\right)^2  \right)$ . Generically it spans a 3d subspace of the 2-forms of each horizontal space and thus allows us to defines a conformal metric (the associated Urbantke metric cf section \ref{section: Chiral formulation of gravity}) \emph{on each horizontal tangent spaces}. Let's see how it works explicitly:\\ Define
\begin{equation}
B^{A'B'} \coloneqq \frac{\p^{A'}\p^{B'}}{\left( \pp\right)^2} d_a\gtb + \frac{\ph^{A'}\ph^{B'}}{\left( \pp\right)^2} d_a\gt + \frac{\p^{(A'}\ph^{B')}}{\pp} \left( d_a\right)^2
\end{equation}
and
\begin{equation}
B^i = \gs^i_{A'B'} \;B^{A'B'}.
\end{equation}
This last object should be understood as an $\su2$-valued 2-form. From this we can follow the same procedure as in the first section and construct $\gS$:
\begin{equation}
\gS^i\left( x,\gz\right)= X^{-\frac{1}{2}}{}^{ij}B^j \qquad \gS^{A'B'}=\gs^{A'B'}_i \gS^i
\end{equation}
such that $\gS^i \W \gS^j \propto \gd^{ij}$.
It is associated with a conformal class of metric on each horizontal space $e{}^{AA'}\left( x,\p\right)$, $\gS^{A'B'}= e^{A'}{}_{A}\W e^{B'A}$.

Importantly at this point the tetrad on the horizontal tangent space $e^{AA'}$ varies along the fiber $e^{AA'} = e^{AA'}(x,\p)$.

In the end we define the $\Oc(2)$-valued 2-form on $\PT$: 
\begin{equation}
\gS_{\gt} (x, \pi) = \gS^{A'B'} (x, \p) \p_{A'}\p_{B'} = \theta^0\W \theta^1, \quad \text{with} \; \theta^{A}= e^{AA'}\p_{A'}.
\end{equation}

This construction is not as arbitrary as it might seem at first sight: in the particular case where there is an underlying $\SU(2)$-connection such that $\gt = \p_{A'}\left(d\pi^{A'} + A^{A'}{}_{B'}\pi^{B'} \right)$, it precisely coincides with the construction from proposition \ref{Proposition- Euclidean Twistor Theory Revisited: Almost Hermitian structure}. The connection $a$ on $\Oc(2)$ then coincides with the connection described at the beginning of section \ref{section: Euclidean Twistor Theory Revisited} and the restriction of the triplet  \begin{equation}
\left(d_a\gt,d_a\gtb, \left( d_a\right)^2  \right) 
\end{equation} to the horizontal tangent space then indeed is just 
\begin{equation}
\left(F^{A'B'}\p_{A'}\p_{B'},F^{A'B'}\ph_{A'}\ph_{B'} , F^{A'B'}\p_{A'}\ph_{B'} \right).
\end{equation}
 Note that we did not need to assume the connection to be perfect. It can then be checked that, under such conditions, the twistor action \eqref{Discussion on the would be `Twistor action for Einstein gravity': twistor ansatz 2} coincides with the original  space-time action from proposition \eqref{Proposition- Discussion on the would be `Twistor action for Einstein gravity': Action: generalised CS}.

Therefore we could hope that with this definition for $\gS_{\gt}$, the action \eqref{Discussion on the would be `Twistor action for Einstein gravity': twistor ansatz 2} would describe gravity: all we need are the field equations for $\psi$ to imply the existence of an $\SU(2)$ connection such that $\gt = \p_{A'}\left(d\pi^{A'} + A^{A'}{}_{B'}\pi^{B'} \right)$. 

At this point however, it seems that we are out of luck. Varying \eqref{Discussion on the would be `Twistor action for Einstein gravity': twistor ansatz 2} with respect to $\psi$ we obtain
\begin{equation}
\gt \W d\gt \W \gS_{\gt} + \eps \gt \W \gS_{\gt} \W B^{A'B'} \p_{A'}\p_{B'}=0.
\end{equation}

For simplicity let's consider the case $\eps=0$. Then the action in proposition \ref{Proposition- Discussion on the would be `Twistor action for Einstein gravity': Action: generalised CS} is an action for self-dual gravity and we thus would like to interpret, 
\begin{equation}\label{Discussion on the would be `Twistor action for Einstein gravity': Twistor self dual field equations}
	\gt \W d\gt \W \gS_{\gt}=0
\end{equation}
as implying the integrability of some almost complex structure and/or as the perfectness of some $\SU(2)$ connection arising on the way.

However, on the one hand, due to the important non linearities involved in constructing $\gS_{\gt}$, it seems very complicated to interpret this field equations in terms of the almost complex structure from section \ref{ssection: the NLGT revisited}. On the other hand, one could be tempted to consider as another almost complex structure: the one that makes $\gt \W \gS_{\gt}$ a $(3,0)$-form, then the field equations \eqref{Discussion on the would be `Twistor action for Einstein gravity': Twistor self dual field equations} just read $d\gt \big|_{0,2}=0$.

At this point, to obtain self-dual gravity, it would be enough to be able to conclude that there exists a $\SU(2)$ connection such that $\gt = \p_{A'} \left(d\p^{A'} + A^{A'}{}_{B'} \p^{B'}\right)$.

This is however not the case: generically $\gt$ can be written $\gt = d\gz + A_{\mu} dx^{\mu} $ with $A_{\mu} \in \Gamma(\Oc(2))$. From this is follows that
\begin{equation}
 \pab_{\gzb} \id d\gt =0 \qquad \Leftrightarrow \qquad \pab_{\gzb}\left(A_{\mu}\right)dx^{\mu} = 0 
\end{equation}
would indeed imply that $A_{\mu}= A^{A'B'}\p_{A'}\p_{B'}$. On the other hand 
\begin{equation}
\pab_{\gzb} \id \left(d\gt \big|_{0,2}\right) =0 \qquad \Leftrightarrow \qquad \pab_{\gzb}\left(A_{\mu}\right) dx^{\mu}\big|_{0,1} = 0
\end{equation}
are just not enough field equations to conclude that $\pab_{\gzb}\left(A_{\mu}\right)=0$. In this last case one indeed misses one half of the necessary field equations:
\begin{equation}
 \pab_{\gzb} \id \left(d\gt \big|_{1,1}\right) =0.
\end{equation} 

In principle, this missing set of equations could be implemented as a constraint in our twistor ansatz \eqref{Discussion on the would be `Twistor action for Einstein gravity': twistor ansatz 2}: this would at last gives a twistor action for gravity. However, on top of definitely spoiling any remaining geometric aesthetics, it would also add another layer of complexity to our already complicated construction, making it more than unlikely to be useful.

\section*{Conclusion}\pdfbookmark[section]{Conclusion}{Section: Conclusion}

In this paper we reviewed chiral formulations of gravity, in particular the pure connection formulation, and showed that they nicely interact with twistor theory: from a definite $\SU(2)$-connection only on a 4d manifold $M$ one can construct an almost hermitian structure on the associated twistor space $\PT(M)$, cf proposition \ref{Proposition- Euclidean Twistor Theory Revisited: Almost Hermitian structure}. The holomorphicity of a certain 1-form $\gt$ is then equivalent to the requirement that the connection is the self-dual part of a anti-self-dual space-time, cf proposition \ref{Proposition- Euclidean Twistor Theory Revisited: Integrability of A}. On the other hand such a 1-form $\gt$ on a 6d manifold $\PTc$ defines an almost complex structure $\J_{\gt}$ and a 4d manifold $\CP^1 \inj \PTc \to M$. Finally, integrability of this almost complex structure allows to define a $\SU(2)$-connection such that it is the self-dual connection of an anti-self-dual metric, cf proposition \ref{Proposition- Euclidean Twistor Theory Revisited: Non Linear graviton}.

Chiral formulations of gravity come in the form of many different Lagrangians that we reviewed in appendix \ref{section: Appdx A Review of Chiral Lagrangians for Gravity} and together with our description of twistor theory in terms of connections they suggest different ansätze for a twistor action for gravity. We described some of them in section \ref{section: Discussion on the would be `Twistor action for Einstein gravity'}. As to now however, none of them seems to lead to a useful twistor action and the chase for such an action (if only it exists!) is still open. 

\section*{Acknowledgments}
The author is most grateful to Kirill Krasnov for numerous discussions on the topics touched on in this paper. He is also pleased to acknowledge important remarks and suggestions from Lionel Mason on the materials described in section 2 and 3. 

\appendix
\section{A Review of Chiral Lagrangians for Gravity} \label{section: Appdx A Review of Chiral Lagrangians for Gravity}

\subsection{The Plebanski action, $S[A,\Psi, B]$: }\label{section Appdx, ss: The Plebanski Action}

The Plebanski Action for General Relativity is
\begin{equation}
	S[\bdA,\bdB,\Psi] = \int B^i \W F^i - \left(\Psi^{ij} + \frac{\gL}{3}\gd^{ij}\right) \frac{B^i \W B^j}{2}
\end{equation}
see \cite{Plebanski:1977zz}, \cite{Capovilla:1991qb} for the original references.\\
It is not a very economical action as is contains many fields: a $\SU(2)$-connection $\bdA$ (which does not need to be a definite connection at this stage), a $su(2)$-valued 2-form $\bdB$ (again we do not need to require this triplet of 2-forms to be a definite triplet) and a symmetric \emph{traceless} field: $\Psi^{(ij)} = \Psi^{ij}$.

Varying the action with respect to $\Psi^{ij}$ we get
\begin{equation}\label{Plebanski Eq: dpsi}
	\frac{\gd S}{\gd \Psi^{ij}}=0 \qquad \Leftrightarrow \qquad B^i \W B^j \propto \gd^{ij} d^4x.
\end{equation}
Thus the triplet of 2-form $\{B^i\}_{i \in 1..3}$ is definite: It implies that there is a unique Euclidean, invertible, conformal metric $\tilde{g}_B$ such that the triplet is a basis of self-dual 2-forms. What's more the field equations implies that $\left\{B^i\right\}_{i\in 1,2,3}$ is an orthonormal basis for the wedge product. This means that, up to a $\SU(2)$ action, the $B$'s can in fact be constructed from the conformal metric $\tilde{g}_{B}$ as in eq \eqref{Chiral Formulations of GR - Fundations: Sigma def (tetrad)}.
We can pick up a representative $g_B$ in the conformal class $\tilde{g}_{B}$ by requiring the volume form to be $vol_g = B^i \W B^i$.

Varying the action with respect to $\bdA$ we then have
\begin{equation}\label{Plebanski Eq: dA}
	\frac{\gd S}{\gd A^i}=0 \qquad \Leftrightarrow \qquad d_A B^i =0.
\end{equation}
This equation, together with \eqref{Plebanski Eq: dpsi}, implies that $\bdA$ is the self-dual part of the Levi-Civita connection associated  with the metric $g_B$.

Varying the action with respect to $\bdB$ we find,
\begin{equation}\label{Plebanski Eq: dB}
	\frac{\gd S}{\gd B^i}=0 \qquad \Leftrightarrow \qquad F^i = \left(\Psi^{ij} + \frac{\gL}{2}\gd^{ij}\right) B^j.
\end{equation}
which is now just the chiral version of Einstein equations, cf \eqref{Chiral Formulations of GR - Fundations: Chiral Einstein Equations 2}.

\subsection{The self-dual Palatini action $S[A,e]$}

The self-dual Palatini action (or Ashtekar action) really is the covariant side of the canonical description of gravity in terms of Ashtekar variables (see \cite{Jacobson:1988yy}, \cite{Peldan:1993hi} for a precise derivation of the constraints). 
On can obtain this action by varying $\Psi$ in the Plebanski action and solving the associated `simplicity constraints' $B^i \W B^j \propto \gd^{ij}$. As we already stated this can be done by introducing the unique tetrad ${e^I}_{I\in 0..3}$ such that
\begin{align}
	&\left\{B^i= \gS^i_{IJ} \frac{e^I \W e^J}{2} = -e^0 \W e^i - \frac{\eps^{ijk}}{2} e^j \W e^k  \right\}_{i\in 1,2,3} .
\end{align}
The resulting action is
\begin{equation}
	S[\bdA,e] = \int B^i(e) \W F^i + \gL\;\frac{\eps_{IJKL}}{4} e^I \W e^J \W e^K \W e^L.
\end{equation}
Varying the action with respect to the connection we get the compatibility equation $d_{\bdA} \bdB=0$. Using the decomposition \eqref{Appdx: Decomposition of the Riemann tensor}, one sees that integrating out $\bdA$ from the self-dual Palatini action gives the usual Einstein Hilbert.

\subsection{Intermediate actions of the type $S[A,\Psi]$ : }\label{section Appdx, ss: Intermediate action} 

Let's start again from the Plebanski action, we saw that integrating out $\Psi$ lead to the self-dual Palatini action. Instead, we now want to eliminate the $\bdB$ field: making use of equations \eqref{Plebanski Eq: dB}, the resulting action is:
\begin{equation}\label{Intermediate action I}
	S[\bdA,\Psi] = \int \left(\left(\Psi +\frac{\gL}{3} \gd \right)^{-1}\right)^{ij} F^i\W F^j
\end{equation}
We let the reader check that the this action yields field equations for GR. As for all chiral formulations of GR the mechanism comes done to constructing a metric with Urbantke trick \eqref{Chiral Formulations of GR - Fundations: Urbantke metric}, and the checking the compatibility relations \eqref{Chiral Formulations of GR - Fundations: Sigma/D compatibility} $d_{\bdA} \bdgS =0$ (here $\gS^i = \left(\left(\Psi +\frac{\gL}{3} \gd \right)^{-1}\right)^{ij} F^j$).

An interesting variant of the action \eqref{Intermediate action I} is obtained by expanding the inverse matrix in power of $\Psi$,
\begin{equation}
	S[\bdA, \Psi] = \frac{3}{\gL}\int F^i \W F^i +   \sum_{k=1}^{\infty} \left(\frac{3}{\gL}\right)^k \; \left(\Psi^k\right)^{ij} F^i \W F^j
\end{equation}
By subtracting the topological term $F^i \W F^i$ and multiplying by $\left(\frac{\gL}{3}\right)^2$ we obtain:
\begin{equation}\label{Intermediate action III}
	S[\bdA, \Psi] = \int \Psi^{ij}\; F^i\W F^j + \sum_{k=0}^{\infty} \left(\frac{3}{\gL}\right)^{k+1} \; \left(\Psi^{k+2}\right)^{ij} F^i \W F^j
\end{equation}
This action, which does not seem to have attracted much attention up to now, has the following interesting interpretation: in the limit where $\gL$ goes to infinity we recover an action for anti-self-dual gravity. For a finite $\gL$ however this action describe full GR as an interacting theory around the anti-self-dual background with the cosmological constant playing the role of coupling constant. This parallels the Chalmers-Siegel action for Yang-Mills. 

\subsection{Intermediate actions of the type $S[A,B]$}

Before we come to the pure connection action let's consider the action for gravity of the form $S[A,B]$. It has already been described in \cite{Herfray:2015rja} and cannot be obtained from the Plebanski action (see however \cite{Celada:2016iah} for a derivation from a more complicated Lagrangian):
\begin{equation}
S\left[\bdA, \bdB\right] = \int F^i \W F^i + \left( Tr \sqrt{B \W B} \right)^2 + \frac{\eps}{2} B^i\W B^i
\end{equation}

As described in \cite{Herfray:2015rja}, for $\eps=0$ one recovers again anti-self-dual gravity while for $\eps\neq0$ it describes full gravity. 

\subsection{Pure connection action $S[A]$:} \label{section Appdx, ss: Pure Connection action}

We now come to the pure connection action, starting again from \eqref{Intermediate action I} and integrating $\Psi^{ij}$ one obtains:
\begin{equation}
	S[\bdA] = \int \left(Tr\sqrt{ F \W F}\right)^2.
\end{equation} 
See proposition \ref{Proposition- Definite Connections and Gravity: Pure connection equation} for a discussion on the pure connection Einstein field equations and \cite{Krasnov:2011pp} for the original reference.

\subsection{The background independent Chalmer-Siegel action for $\SU(2)$}

We end up this review of chiral Lagrangians with the action described in proposition \ref{Proposition- Discussion on the would be `Twistor action for Einstein gravity': Action: generalised CS} and that was previously unknown in the literature:

As we already explained in section \ref{ssection: A new action for Gravity as a background invariant generalisation of the Chalmers-Siegel action.}, using the property of definite $\SU(2)$- connections to parametrize space-time metric, one can turn the Chalmer-Siegel action for $\SU(2)$ connection:
\begin{equation}
	S\left[\bdA, \bdB\right] = \int_M  B^i\W F^i - \frac{\eps}{2} B^i \W B^i 
\end{equation}
into a background independent action:
\begin{equation}
S[\bdA, B^{ij}] = \int B^{ij} \gS_A^i\W F^j  - \frac{\eps}{2} B^{ij}B_{ik}\; \gS_A^j\W \gS_A^k.
\end{equation}

As described in proposition \ref{Proposition- Discussion on the would be `Twistor action for Einstein gravity': Action: generalised CS}, for $\eps\neq0$ this action describes gravity at the condition of $B^{ij}$ to be traceless. In the case where $\eps=0$ this action describes anti-self-dual gravity.

\section{Decomposition of the Riemann Curvature Tensor in Coordinates}\label{section: Appdx Decomposition of the Curvature}

In this appendix we prove, using coordinates, the different claims made in the first part of section \ref{section: Chiral formulation of gravity}.

In this appendix we use freely the isomorphism $\so(4)\simeq \gO^2$ to represent elements of $\so(4, \R)$ as 2-forms. I.e, we pick up a basis of 1-forms, $\left\{e^I\right\}_{I\in \{0...3\}}$ compatible with the metric, $ds^2= e^I \otimes e^I$, and write  for $\boldsymbol{b} \in \su(2)$ as $ \boldsymbol{b} = b_{IJ} \frac{e^I \W e^J}{2} $, with abuse of notation. The metric allows to raise and lower $I,J,K...$ indices. With this notations, the Lie bracket reads,
\begin{equation}
\boldsymbol{a},\boldsymbol{b}\in \so(4), \qquad \left[\boldsymbol{a},\boldsymbol{b}\right] = \left( a_{I}{}^{K} b_{KJ} - b_{I}{}^{K} a_{KJ} \right)\frac{e^I \W e^J}{2}.
\end{equation}
Then for any $\boldsymbol{b} \in \su(2)$ the decomposition $\so(4) = \su(2)\oplus \su(2)$ reads:
\begin{equation}\label{Appdx: 2-form decomposition}
	b_{IJ} \frac{e^I\W e^J}{2}= B^i \frac{\gS^i}{2} + \widetilde{B}^i \frac{\gSt^i}{2}. 
\end{equation} 
Where the sigma tensors coincide with the one described in section \ref{sssection: Two useful tensors}. In terms of the tetrad they take the explicit form:
	\begin{align}\label{Appdx: Sigma def (tetrad)}
		&\left\{\gS^i = -e^0 \W e^i - \frac{\eps^{ijk}}{2} e^j \W e^k\right\}_{i\in 1,2,3}, 
		&\left\{\gSt^i = e^0 \W e^i - \frac{\eps^{ijk}}{2} e^j \W e^k\right\}_{i\in 1,2,3}.
	\end{align}
	
	They form a basis of self-dual and anti-self-dual 2-forms respectively. This basis is orthogonal for the wedge product:
	\begin{equation}\label{Appdx: Sigma orthogonality}
		\gS^i \W \gS^j = -\gSt^i \W \gSt^j = 2 \gd^{ij} e^0\W e^1 \W e^2 \W e^3, \qquad \gS^i \W \gSt^j = 0.
	\end{equation}
As was already stated in the main part of this paper, the decomposition of Lie algebra $\so(4)=\su(2)\oplus \su(2)$  corresponds to the decomposition $\gO^2 = \gO^2_+ \oplus \gO^2_-$ of 2-forms:
\begin{equation}
	\left[\gS^i ,\gS^j \right] = 2\eps^{ijk} \gS^k, \quad \left[\gSt^i ,\gSt^j \right] = 2\eps^{ijk} \gSt^k, \quad \left[\gS^i ,\gSt^j \right] = 0.
\end{equation}

In what follows we will make an important use of the tensors, $\gS^i_{IJ}$, $\gSt^i_{IJ}$ defined by $\gS^i = \gS^i_{IJ} \frac{e^I\W e^J}{2}$, $\gSt^i = \gSt^i_{IJ} \frac{e^I\W e^J}{2}$. They verify the algebra,
	\begin{equation}\label{Appdx: Sigma algebra}
		\begin{array}{ccc}
			\gS^i{}_{IK} \gS^j{}^K{}_J = -\gd^{ij} g_{IJ} +\eps^{ijk}\gS^k_{IJ}, \qquad \gSt^i{}_{IK} \gSt^j{}^K{}_J =-\gd^{ij} g_{IJ} +\eps^{ijk}\gSt^k_{IJ} ,\\ \\ \gS^{i}{}_{IK} \gSt^{j}{}^K{}_J = s^{ij}_{IJ}.
		\end{array}
	\end{equation}
Where $s^{ij}_{IJ}$ is a tensor with the following symmetries:
\begin{equation}
	s^{ij}_{[IJ]}=0, \quad s^{[ij]}_{IJ}=0.
\end{equation}
Note that (anti)-self-duality explicitly reads
	\begin{equation}\label{Appdx: Sigma self-duality}
		\gS^i{}^{IJ} = \frac{\eps^{IJKL}}{2} \gS^i_{KL},\qquad \qquad \gSt^i{}^{IJ} = -\frac{\eps^{IJKL}}{2} \gSt^i_{KL}.
	\end{equation}

\paragraph{Decomposition of the Curvature tensor in coordinates} \mbox{} \\
Consider a 4d Riemannian manifold $\{g, M\}$, $\{e^I\}_{I\in 0..4}$ an orthonormal co-frame. The Levi-Civita connection, $\nabla $, then naturally is a $SO(4)$-connection. We will write its potential 1-form $\boldsymbol{a}$ and curvature 2-form $\boldsymbol{f}$ as
\begin{equation}
	a^I{}_J = a^{I}{}_{J\;K} e^K, \qquad f^I{}_J = d a^I{}_J + a^I{}_K \W a^K{}_J = f^I{}_J{}_{KL} \frac{e^K \W e^L}{2}.
\end{equation}
Note that the Riemann curvature $\bdf$ here is a $\so(4)$-valued 2-form.

Now we can use the decomposition $\so(4)= \su(2) \oplus \su(2)$, concretely realised as \eqref{Appdx: 2-form decomposition}, to define the chiral connections $\left(D,\Dt \right)$ with potential $\left(A,\At \right)$ as
\begin{equation}\label{Appdx: LeviCivita split}
	a^I{}_J = A^i \frac{\gS^i{}^I{}_J}{2} + \At^i \frac{\gSt^i{}^I{}_J}{2}.
\end{equation}
These connections naturally are $\SU(2)$-connections.  In section \ref{section: Chiral formulation of gravity} we stated that these connections are compatible with $\gS^i$, $\gSt^i$ in the following sense:
\begin{equation}\label{Appdx: Sigma/A compatibility}
	d_A \gS^i = 0, \qquad d_{\At} \gSt^i=0.
\end{equation}
We can prove this by a direct computation:
\begin{proof}
	
	\begin{align*}
		d_A \gS^i &= \frac{1}{2} d_A \left( \gS^i_{IJ} e^I \W e^J \right)\\
		&=\frac{1}{2} d_A \left( \gS^i_{IJ}\right) e^I \W e^J\\ 
		& = \frac{e^I \W e^J}{2} \W  \left( \eps^{ijk} A^j \gS^k_{IJ} - 2\gS^i_{IK} a^K{}_J  \right) \\
		&= \frac{e^I \W e^J}{2} \W  \left( \eps^{ijk} A^j \gS^k_{IJ} - 2A^j \gS^i_{IK}  \frac{\gS^j{}^K{}_J}{2} -2 \At^j \gS^i_{IK}\frac{\gSt^j{}^K{}_J}{2}  \right)\\
		&=0
	\end{align*}
	where in step 1 we used the torsion freeness of $a$ (i.e $d_a e^I =0$), step 3 is just the decomposition of the Levi Civita connection into its chiral parts, (i.e, eq\eqref{Appdx: LeviCivita split}) and at step 4 we made use the algebra \eqref{Appdx: Sigma algebra}.
\end{proof}

As already stated in the main body of this paper, the relations \eqref{Appdx: Sigma/A compatibility} can be used as an alternative way of defining $\bdA$(resp $\At$) as the unique $\SU(2)$-connection compatible with $\gS^i$(resp $\gSt^i$). 

\begin{proof}
	
	Let us suppose that $\bdA$ and $\bdA' = \bdA +\bdM$ are are both $\SU(2)$-connections compatible with $\gS^i$. It follows that
	\begin{equation}
		d_{A'} \gS^i - d_A \gS^i = \eps^{ijk} M^j \gS^k =0,
	\end{equation}
	or equivalently, by making use of the self-duality of $\gS$,
	\begin{equation}\label{Appdx: proof A unicity (1)}
		M_{\nu}^{[i} \gS^{j]}{}^{\mu\nu}=0.
	\end{equation}
	By multiplying  this expression by another sigma symbol we have
	\begin{align}
		0&= \eps^{ijk} M^j_{\nu} \gS^k{}^{\mu\nu} \gS^l_{\mu\gr}\\
		&= \eps^{ijk} \eps^{klm} M^j_{\nu} \gS^m{}^{\nu}{}_{\gr} + \eps^{ijl} M^j_{\gr} \\
		&=\gd^{il} M^k_{\nu} \gS^k{}^{nu}{}_{\gr} -M^l_{\nu}\gS^i{}^{\nu}{}_{\gr} +  \eps^{ijl} M^j_{\gr} 
	\end{align}
	where we made use of the algebra \eqref{Appdx: Sigma algebra} and the identity $\eps^{abm} \eps^{ijm} = \gd^{ai}\gd^{bj} - \gd^{aj}\gd^{bi}$. Anti-symmetrising this last expression  in the $i,l$ indices and making use of \eqref{Appdx: proof A unicity (1)} we obtain
	\begin{equation}
		0=-M^{[l}_{\nu}\gS^{i]}{}^{\nu}{}_{\gr}+ \eps^{ilj} M^j_{\gr} = \eps^{ilj} M^j_{\gr}.
	\end{equation}
	Which conclude the proof that there is a unique connection satisfying \eqref{Appdx: Sigma/A compatibility}.
\end{proof}

In complete parallel with \eqref{Appdx: LeviCivita split} we define the `self-dual part of the Curvature' $F$ and the `anti-self-dual part of the Curvature' $\Ft$ as
\begin{equation}\label{Curvature 2-form split}
	f^I{}_J = F^i \frac{\gS^i{}^I{}_J}{2} + \Ft^i \frac{\gSt^i{}^I{}_J}{2},
\end{equation}
and these are naturally $\su(2)$-valued 2-forms.
In fact we have,
\begin{equation}
	F^i = dA^i +\frac{\eps^{ijk}}{2} A^j \W A^k, \qquad \Ft^i = d\At^i +\frac{\eps^{ijk}}{2} \At^j \W \At^k,
\end{equation}
as can be seen using the algebra \eqref{Appdx: Sigma algebra}. I.e, the (anti-)self-dual part of the curvature is the curvature of the (anti-)self-dual connection.

Now $F^i$, $\Ft^i$ being ($\su(2)$-valued) 2-forms, we can decompose them into self-dual and anti-self-dual pieces: 
\begin{equation}
	F^i = F^{ij} \gS^j + G^{ij} \gSt^j, \qquad \Ft^i = \Gt^{ij} \gS^j + \Ft^{ij} \gSt^j.
\end{equation}
This is just another way of writing the bloc decomposition \eqref{Chiral Formulations of GR - Fundations: F, Ft decomposition}.
The Riemann curvature now reads
\begin{equation}
	f^I{}_J = \frac{1}{2}\left(  F^{ij}\; \gS^j \; \gS^i{}^I{}_J + G^{ij} \; \gS^j \;\gSt^i{}^I{}_J +  \Gt^{ij} \;\gSt^j \;\gS^i{}^I{}_J + \Ft^{ij} \;\gSt^j \; \gSt^i{}^I{}_J \right).
\end{equation}
Again, this is just another version of the bloc decomposition \eqref{Chiral Formulations of GR - Fundations: Riemann decomposition 1}.
To get the final form of this decomposition we write 
\begin{equation}
	F^{ij} = \frac{\gL}{3} \gd^{ij} +\Psi^{ij}, \qquad \Ft^{ij} = \frac{\gLt}{3} \gd^{ij} +\Psit^{ij},
\end{equation}
with $\Psi$, $\Psit$ some traceless tensors and $\gL = tr F$, $\gLt =tr\Ft$. Finally, we can write:

	\begin{align}\label{Appdx: Decomposition of the Riemann tensor}
		f^I{}_J &= \underbrace{\frac{\gLt}{3} \;\gSt^i \; \frac{\gSt^i{}^I{}_J}{2} + \frac{\gL}{3} \;\gS^i \; \frac{\gS^i{}^I{}_J}{2}}_{\text{Scalar Part}}    \nonumber\\
		&+ \underbrace{ \frac{1}{2}\left(G^{ij} \; \gS^j \;\gSt^i{}^I{}_J +  \Gt^{ij} \;\gSt^j \;\gS^i{}^I{}_J+\Psi^{[ij]} \;\gS^j \; \gS^i{}^I{}_J + \Psit^{[ij]} \;\gSt^j \; \gSt^i{}^I{}_J\right)}_{\text{Ricci Part}} \nonumber\\& + \underbrace{\Psi^{(ij)} \;\gS^j \; \frac{\gS^i{}^I{}_J}{2} + \Psit^{(ij)} \;\gSt^j \; \frac{\gSt^i{}^I{}_J}{2}}_{\text{Weyl Part}} 
	\end{align} 
	This is just a hands-on way of rewriting the bloc decomposition \eqref{Chiral Formulations of GR - Fundations: Riemann decomposition 1}. One can identify the following elementary brick of the Riemann tensor:
\begin{equation}\label{Appdx: Decomposition of the Riemann tensor 2}
\begin{array}{ll}
W_{IJKL}= \frac{1}{2}\Psi^{(ij)} \;\gS^i_{IJ} \; \gS^j{}_{KL} \quad  &\text{is the self-dual part}\\ & \text{of the Weyl~tensor}, \\ \\
\Wt_{IJKL} = \frac{1}{2}\Psit^{(ij)} \;\gSt^i_{IJ} \; \gSt^j{}_{KL} \quad  &\text{is the anti-self-dual part } \\ &  \text{of the Weyl~tensor,}\\ \\
R = 2\gL +2\gLt \quad &\text{is the Scalar curvature,} \\ \\
\text{and the traceless Ricci tensor is},& \\ \\ R^{I}{}_{J} = \underbrace{\frac{1}{2} \left(G^{ji}+\Gt^{ij} \right) \gS^i{}^{KI}\; \gSt^j{}_{KJ}}_{\text{symetric traceless Ricci}} & \underbrace{-\frac{1}{2}\left(\Psi^{[ij]} \;e^{ijk}\; \gS^k{}^I{}_{J} \; +\Psit^{[ij]} \;e^{ijk}\; \gSt^k{}^I{}_{J}\right)}_{\text{anti-symetric Ricci}}   \\ \\

\end{array}
\end{equation}
These can be related to the usual definitions
\begin{equation}
	R = f^{KL}{}_{KL}, \qquad R^I{}_J = f^{KI}{}_{KJ}-\frac{1}{4}R\;\gd^I{}_{J} \,,\quad \text{with}\; F^I{}_J = F^I{}_{JKL} \frac{e^K \W e^L}{2}
\end{equation}
by contracting indices in \eqref{Appdx: Decomposition of the Riemann tensor} and using the algebra \eqref{Appdx: Sigma algebra}.

Consequently Einstein equations are equivalent to
\begin{equation}
	\Psi^{[ij]}=0,\quad \Psit^{[ij]}=0, \quad G^{ij} + \Gt^{ji}=0.
\end{equation}

As stated in the main body of this paper the torsion freeness of the Levi-Civita connection implies that the Riemann tensor has some internal symmetries (Bianchi I) that lead to further simplifications:
\begin{equation}
	f_{IJKL} = f_{KLIJ} \quad \Rightarrow \quad \Psi^{ij}=\Psi^{(ij)},\quad \Psit^{ij}=\Psit^{(ij)}, \quad \Gt^{ij}=G^{ji}.
\end{equation}
\begin{equation}
	f_{I[JKL]}=0 \quad \Leftrightarrow\quad  f_{NIKL} \eps^{NJKL}=0 \quad \Rightarrow \quad \gL = \gLt.
\end{equation}
The second relation follows from using the (anti)-self duality (see eq \ref{Appdx: Sigma self-duality}) of the sigma tensors.

With those symmetries, Einstein equations
\begin{equation}
	R_{IJ} = \gL g_{IJ}, 
\end{equation}
are equivalent to
\begin{equation}\label{Appdx: Chiral Einstein equations}
	F^i = \left( \Psi^{ij} + \frac{\gL}{3} \gd^{ij} \right) \gS^{j}
\end{equation} 
(i.e  $G=0$ ) and we therefore need only one half of the Riemann tensor to state them.

\section{Spinor conventions} \label{section: Appdx Spinor conventions} \mbox{}

\indent We convert $\su(2)$ lie algebra indices into spinor notations according to the rule:
\begin{align*} 
	V = V^i \gs_i \in \su(2), \quad V^i=\{x,y,z\} & \Leftrightarrow &  V^i \gs_i^{A}{}_{B} = V^{A}{}_{B} = \frac{1}{2i}\Mtx{z & x-iy \\
		x+iy & -z} \in \su(2).
\end{align*}
Such that $\left[\gs^i, \gs^j \right] = \eps^{ijk} \gs^k$. Latin indices are raised and lowered with the flat metric $\gd^{ij}$, spinor indices are raised and lowered as usual using the antisymmetric $\eps^{A'B'}$ as explained in the beginning of section 2. We go from one type of indices to the other as follows:
\begin{equation}
V^{A'B'} = V^i \,\gsb_i^{A'B'} \qquad \Leftrightarrow \qquad V^i = 2\gsb^i_{A'B'} V^{A'B'}.
\end{equation}

	Note however that, as a convention, we use $\gs_i^{AB}$ to convert "spatial indices" $i\in \{1,2,3\}$ into `unprimed spinor indices' and $\gsb_i^{A'B'}$, its complex conjugate, to convert "spatial indices" into "primed spinor indices".

On the other hand, to convert space-time indices into spinor ones we use the convention:
\begin{align} 
	V^I e_I^{AA'} = \frac{1}{i\sqrt{2}}\Mtx{-it+z & x-iy \\  x+iy & -it-z}, \qquad  V^I = \{t,x,y,z\}.
\end{align}
With this conventions, a direct computation shows that
\begin{equation}
\gS^i = 2 \gsb^i_{A'B'}\; \frac{1}{2} e^{A'C} \W e^{B'}{}_{C} = -e^0 \W e^i - \eps^{ijk} e^i \W e^k, \qquad i,j,k \in \{1,2,3\}.
\end{equation}
i.e 
\begin{equation}
\gS^{A'B'} = \gsb_i^{A'B'} \gS^{i} =  \frac{1}{2} e^{A'C} \W e^{B'}{}_{C}.
\end{equation}

\bibliography{Biblio}

\begin{thebibliography}{10}

\bibitem{Plebanski:1977zz}
Jerzy~F. Plebanski.
\newblock {On the separation of Einsteinian substructures}.
\newblock {\em J. Math. Phys.}, 18:2511--2520, 1977.

\bibitem{Jacobson:1988yy}
Ted Jacobson and Lee Smolin.
\newblock {Covariant Action for Ashtekar's Form of Canonical Gravity}.
\newblock {\em Class. Quant. Grav.}, 5:583, 1988.

\bibitem{Krasnov:2011pp}
Kirill Krasnov.
\newblock {Pure Connection Action Principle for General Relativity}.
\newblock {\em Phys.Rev.Lett.}, 106:251103, 2011.

\bibitem{Ashtekar:1986yd}
A.~Ashtekar.
\newblock {New Variables for Classical and Quantum Gravity}.
\newblock {\em Phys. Rev. Lett.}, 57:2244--2247, 1986.

\bibitem{Capovilla:1991qb}
R.~Capovilla, T.~Jacobson, J.~Dell, and L.~J. Mason.
\newblock {Selfdual two forms and gravity}.
\newblock {\em Class. Quant. Grav.}, 8:41--57, 1991.

\bibitem{Capovilla:1991kx}
R.~Capovilla, T.~Jacobson, and J.~Dell.
\newblock {A Pure spin connection formulation of gravity}.
\newblock {\em Class. Quant. Grav.}, 8:59--73, 1991.

\bibitem{Penrose:1999cw}
Roger Penrose.
\newblock {The Central programme of twistor theory}.
\newblock {\em Chaos Solitons Fractals}, 10:581--611, 1999.

\bibitem{Penrose:1976js}
R.~Penrose.
\newblock {Nonlinear Gravitons and Curved Twistor Theory}.
\newblock {\em Gen.Rel.Grav.}, 7:31--52, 1976.

\bibitem{Ward:1980am}
R.S. Ward.
\newblock {Self-dual space-times with cosmological constant}.
\newblock {\em Commun.Math.Phys.}, 78:1--17, 1980.

\bibitem{Livine:2011vk}
Etera~R. Livine, Simone Speziale, and Johannes Tambornino.
\newblock {Twistor Networks and Covariant Twisted Geometries}.
\newblock {\em Phys. Rev.}, D85:064002, 2012.

\bibitem{Speziale:2012nu}
Simone Speziale and Wolfgang~M. Wieland.
\newblock {The twistorial structure of loop-gravity transition amplitudes}.
\newblock {\em Phys. Rev.}, D86:124023, 2012.

\bibitem{Mason:2008jy}
L.J. Mason and David Skinner.
\newblock {Gravity, Twistors and the MHV Formalism}.
\newblock {\em Commun.Math.Phys.}, 294:827--862, 2010.

\bibitem{Atiyah:1978wi}
M.F. Atiyah, Nigel~J. Hitchin, and I.M. Singer.
\newblock {Selfduality in Four-Dimensional Riemannian Geometry}.
\newblock {\em Proc.Roy.Soc.Lond.}, A362:425--461, 1978.

\bibitem{Fine:2008}
Joel Fine and Dmitri Panov.
\newblock {Symplectic Calabi–Yau manifolds, minimal surfaces and the
  hyperbolic geometry of the conifold}.
\newblock 2008.

\bibitem{Fine:2011}
Joel Fine.
\newblock {A Gauge Theoretic Approach to anti-self-dual Einstein Equations}.
\newblock 2011.

\bibitem{Capovilla:1990qi}
R.~Capovilla, T.~Jacobson, and J.~Dell.
\newblock {GRAVITATIONAL INSTANTONS AS SU(2) GAUGE FIELDS}.
\newblock {\em Class. Quant. Grav.}, 7:L1--L3, 1990.

\bibitem{Mason05}
L.J. Mason.
\newblock {Twistor actions for non-self-dual fields: A Derivation of
  twistor-string theory}.
\newblock {\em JHEP}, 0510:009, 2005.

\bibitem{Wolf:2007tx}
Martin Wolf.
\newblock {Self-Dual Supergravity and Twistor Theory}.
\newblock {\em Class.Quant.Grav.}, 24:6287--6328, 2007.

\bibitem{Adamo:2013tja}
Tim Adamo and Lionel Mason.
\newblock {Conformal and Einstein gravity from twistor actions}.
\newblock {\em Class.Quant.Grav.}, 31(4):045014, 2014.

\bibitem{Mason&Wolf09}
L.J. Mason and Martin Wolf.
\newblock {Twistor Actions for Self-Dual Supergravities}.
\newblock {\em Commun.Math.Phys.}, 288:97--123, 2009.

\bibitem{Adamo:2013cra}
Tim Adamo.
\newblock {Twistor actions for gauge theory and gravity}.
\newblock 2013.

\bibitem{Penrose_vol1}
R.~Penrose and W.~Rindler.
\newblock {\em {Spinors and space-time. Vol 1. Two spinor calculus and
  relativistic fields}}.
\newblock 1985.

\bibitem{Krasnov:2009pu}
Kirill Krasnov.
\newblock {Plebanski Formulation of General Relativity: A Practical
  Introduction}.
\newblock {\em Gen.Rel.Grav.}, 43:1--15, 2011.

\bibitem{Urbantke:1984eb}
H.~Urbantke.
\newblock {On integrability properties of su(2) Yang-Mills Fields}.
\newblock 1984.

\bibitem{Fine:2013qta}
Joel Fine, Kirill Krasnov, and Dmitri Panov.
\newblock {A gauge theoretic approach to Einstein 4-manifolds}.
\newblock {\em New York J.Math.}, 20:293--323, 2014.

\bibitem{Woodhouse85}
N.M.J. Woodhouse.
\newblock {Real methods in twistor theory}.
\newblock {\em Class.Quant.Grav.}, 2:257--291, 1985.

\bibitem{McDuff&Salamon2004}
D.~McDuff, D.~Salamon.
\newblock {\em {J-holomorphic curves and sympleptic topology}}, volume~52 of
  {\em "Colloquium publications"}.
\newblock AMS, 2004.

\bibitem{Jiang08}
Wen Jiang.
\newblock {Aspects of Yang-Mills Theory in Twistor Space}.
\newblock 2008.

\bibitem{Adamo:2011pv}
Tim Adamo, Mathew Bullimore, Lionel Mason, and David Skinner.
\newblock {Scattering Amplitudes and Wilson Loops in Twistor Space}.
\newblock {\em J.Phys.}, A44:454008, 2011.

\bibitem{Chalmers:1996rq}
G.~Chalmers and W.~Siegel.
\newblock {The Selfdual sector of QCD amplitudes}.
\newblock {\em Phys.Rev.}, D54:7628--7633, 1996.

\bibitem{Ward:1977ta}
R.S. Ward.
\newblock {On Selfdual gauge fields}.
\newblock {\em Phys.Lett.}, A61:81--82, 1977.

\bibitem{Peldan:1993hi}
Peter Peldan.
\newblock {Actions for gravity, with generalizations: A Review}.
\newblock {\em Class. Quant. Grav.}, 11:1087--1132, 1994.

\bibitem{Herfray:2015rja}
Yannick Herfray and Kirill Krasnov.
\newblock {New first order Lagrangian for General Relativity}.
\newblock 2015.

\bibitem{Celada:2016iah}
Mariano Celada, Diego González, and Merced Montesinos.
\newblock {Plebanski-like action for general relativity and anti-self-dual
  gravity}.
\newblock {\em Phys. Rev.}, D93(10):104058, 2016.

\end{thebibliography}

\end{document}